# ELECTROSTATIC INTERACTION BETWEEN LONG, RIGID HELICAL MACROMOLECULES AT ALL INTERAXIAL ANGLES

A. A. Kornyshev

*Institute for Theoretical Physics, University of California at Santa Barbara, USA; and Research Center "Jülich", D-52425 Jülich, Germany*

S. Leikin[*]

*Laboratory of Physical and Structural Biology, National Institute of Child Health and Human Development, National Institutes of Health, Bethesda, MD 20892, USA*

We derive formulae for the electrostatic interaction between two long, rigid macromolecules that may have arbitrary surface charge patterns and cross at an arbitrary interaxial angle. We calculate the dependence of the interaction energy on the interaxial angle, on the separation, and on the precise alignment of the charge pattern on one molecule with respect to the other. We focus in particular on molecules with helical charge patterns. We report an exact, explicit expression for the energy of interaction between net-neutral helices in a nonpolar medium as well as an approximate result for charged helices immersed in an electrolyte solution. The latter result becomes exact in the asymptotic limit of large separations. Molecular chirality of helices manifests itself in a torque that tends to twist helices in a certain direction out of parallel alignment and that has a nontrivial behavior at small interaxial angles. We illustrate the theory with the calculation of the torque between layers of idealized, DNA-like double helices in cholesteric aggregates. We propose a mechanism of the observed cholesteric-to-columnar phase transition and suggest an explanation for the observed macroscopic (0.4-5 µm) pitch of the cholesteric phase of *B*-DNA.

## I. INTRODUCTION

Assemblies of long helical macromolecules are common building blocks of living organisms. Bundles of α-helices form domains in many proteins. Bundles of triple-helical collagen form tendons, ligaments, cornea, matrix of skin and bone, and other connective tissue structures. DNA is often stored inside cells and viral capsids in the form of dense aggregates. In other words, many biological interactions involve helices and many complex biological structures consist of helices.

All helices are chiral, i.e. the mirror image of a right-handed helix is left-handed and vice versa. Biological helices manifest their chirality by twisting with respect to each other. Specifically, α-helices form bundles and coiled coils where their long axes cross at a small angle. DNA forms a cholesteric phase that consists of molecular layers. The molecules are parallel within each layer, but their principle axis rotates from layer to layer by a small, constant angle, typically a fraction of a degree [1-6]. Similar assemblies are formed by α-helices in organic solvents [1,7,8], by collagen at low pH [9], and by a variety of other helices. The direction and the amplitude of the twist in cholesteric phases, in helical bundles, and in coiled coils are not random. They are encoded in intermolecular interactions and they are always the same under similar conditions.

---

[*] To whom reprint requests should be addressed. LPSB/NICHD, Bldg. 12A, Rm. 2041, NIH, Bethesda, MD 20892, USA; e-mail leikin@helix.nih.gov; FAX 1-301-496-2172.

The relationship between the twist and the molecular structure is not trivial and still not understood. Some of the puzzles are: (*a*) reversal of the direction of the twist in the cholesteric phase that occurs upon variation in solvent composition even though the handedness of helices remains the same [7]; (*b*) dependence of the amplitude of the twist on the interaxial separation between helices [7,8,10]; (*c*) phase transitions in assemblies of helices, e.g. the transition from a non-chiral line hexatic phase to the chiral cholesteric phase [11,12]; (*d*) anomalously small interaxial angles between helices in the cholesteric phase that could not be justified in terms of structural and dimensional arguments [1-11,13,14]. The solution of these puzzles is likely to be based on understanding properties of chiral interaction potentials between helices at nonzero interaxial angles [15]. Creation of a rigorous theory for such interaction is believed to be one of the biggest challenges of the physics of molecular chirality [15,16].

Electrostatics is a major component of interaction between biological helices since virtually all of them have high density of surface charges. For instance, DNA has an elementary charge per each 1.7 Å of its axial length. As a result, interactions between DNA helices in aggregates are primarily electrostatic. Backbone of an $\alpha$-helix contains a negatively charged carbonyl oxygen and a positively charged amide hydrogen per each 3-4 Å of axial length. The spiral of carbonyls and amides produces an electric field that exponentially decays away from it with the characteristic length ~1 Å. The electrostatic interaction between neighboring $\alpha$-helices in a bundle may be quite strong because of their close contact and their low dielectric constant.

Fifty years ago, a formula for the energy of electrostatic interaction between crossed, homogeneously charged rods in electrolyte solution was derived by Onsager based on the Derjaguin approximation [17]. Twenty five years later, a more general result for the same problem was obtained by Brenner and Parsegian [18] without using the Derjaguin approximation. Recent studies [19,20 21] went beyond the mean field approach and revealed a possibility of attraction at short distances between the rods, induced by ion density fluctuations. However, homogeneously charged rods are not chiral and, therefore, these results tell nothing about the chiral potential. Further progress was impeded by the absence of a mathematical formalism appropriate for chiral helical macromolecules.

To address the effects of molecular chirality, one must take into account the helical structure of a surface charge pattern and its handedness. Electrostatic models for an isolated, charged spiral in an electrolyte solution were recently reported in [22-26]. In particular, the electrostatic potential, counterion distribution around DNA, and the transition between the *B* and *A* forms of DNA in solution were discussed. Electrostatic interactions between several helices that have parallel long axes were calculated in [27]. It was suggested that various details of the helical symmetry of DNA surface charge pattern may be responsible for such observed phenomena as: DNA overwinding from ~10.5 base pairs (bp) per turn in solution to 10 bp/turn in fibers [28], meso- and poly-morphism of DNA in dense aggregates [29], and DNA condensation by counterions [30]. However, none of these results can be directly applied to intermolecular interaction in chiral aggregates (twisted bundles and the cholesteric phase) where molecular axes cross at a non-zero angle.

Development of a theory for electrostatic interactions between long, chiral macromolecules at all interaxial angles is the subject of the present work. First, we derive a result that is valid at arbitrary charge distributions on molecular surfaces. Next, we apply this result to molecules whose surface charge patterns have basic helical symmetries. At the present stage, we incorporate only the most simple, but essential elements of helical structure. We assume that molecules are rigid rods that have helical surface charge patterns. We model a spiral string of point charges by a continuous spiral line charge with the appropriate charge density, helical pitch, and handedness.

Imperfections of molecules and thermal motions are not included into the model. This is not because we believe that the corresponding details are not important, but because one can hope to understand the reality only after thoroughly examining the most simple "ground state" of the system. Once we understand the basics, we will be able to build in details and investigate their role in each specific case.

In general, chiral phenomena are very complex. Each of them deserves a separate, dedicated discussion, because different details are important in each case. Here we focus primarily on laying out the background for the theory. For illustration, we consider interaction between 150 bp *B*-DNA fragments that can be roughly approximated as rigid. We suggest how different observed features of the cholesteric phase formed by such DNA fragments can be qualitatively deduced from properties of the chiral interaction potential between them.

The paper is structured as follows. In Section II, we define the model. In Section III, we present general formulae for the energy of interaction between molecules with helical surface charge patterns and consider simple limiting cases. In Section IV, we discuss interaction between infinitely long helices at small interaxial angles and large separations. In Section V, we show how the results derived for infinitely long helices can be modified and applied to a more realistic case of interaction between two helices of finite length. In Section VI, we analyze main qualitative features expected and observed in the cholesteric phase of 150 bp DNA. All algebra is reported in the Appendix where we derive formulae for the energy of interaction between two molecules with arbitrary surface charge distribution at an arbitrary interaxial angle and apply these formulae to molecules with helical charge patterns.

## II. THE MODEL

We consider interaction between two long, rigid molecules whose axes cross at an arbitrary angle $\psi$ and are separated by the distance $R$ at the point of their closest approach. The configuration and coordinate systems are shown schematically in Fig. 1. The molecules have cylindrical, dielectric inner cores. They are immersed either in an electrolyte solution or in a nonpolar medium. Their intrinsic surface charges form arbitrary, inhomogeneous patterns at surfaces of the inner cores.

We calculate the electric field and the electrostatic interaction energy as a function of $\psi$ and $R$ and of the precise alignment of the charge pattern on one molecule with respect to the other. We start from the interaction Hamiltonian which is the energy at fixed surface charge patterns. Provided that the Hamiltonian for each isolated molecule is known [31], one can then determine the interaction *free energy* by averaging over fluctuations of surface charge densities. In the present work, however, we focus on a more simple case when one can substitute the averaged surface charge pattern into the interaction Hamiltonian to obtain the free energy. This can be done when the energy of *inter*-molecular interaction is much smaller than the *intra*- molecular energy and when the surface charge distributions do not undergo structural transitions, triggered by intermolecular interaction. Such conditions are satisfied in the asymptotic limit of large intermolecular separations or when counterions are strongly bound (chemisorbed) at fixed positions on the surfaces. In these cases, the average surface charge pattern can either be calculated from an adsorption model or it can be approximated on the basis of semi-empirical considerations or computer simulations.

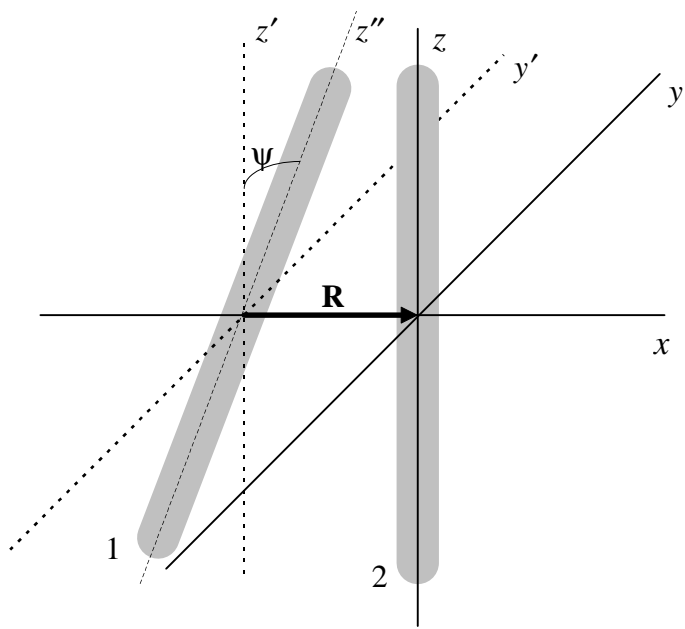

Fig. 1. Configuration of two molecules (1,2) twisted with respect to each other by an *interaxial angle* $\psi$ and shifted by a vector **R** connecting two points of the closest approach on molecular axes. We use three Cartesian coordinate systems: (i) the laboratory frame $(x,y,z)$ whose $z$-axis coincides with the long axis of molecule 2 and $x$-axis goes through the points of closest approach on the axes of molecules 1 and 2; (ii) the shifted laboratory frame $(x',y',z')$ translated by the vector -**R** relative to $(x,y,z)$; and (iii) the molecular frame $(x'',y'',z'')$ whose $z''$-axis coincides with the long axis of molecule 1 and $x''$-axis is coaxial with the $x$-axis of the laboratory frame. The $y''$-axis is not shown to avoid overloading the sketch. In addition to the Cartesian frames, we use two cylindrical molecular frames, each coaxial with the long axis of the corresponding molecule.

Using an average surface charge pattern in the interaction Hamiltonian, we also leave out forces associated with correlated surface charge density fluctuations. Such forces were proposed to play an important role in interaction between homogeneously charged rods [19,20,32-34]. They can affect the character of the interaction and lead to complex non-pairwise-additive forces in bundles of rods [20,34]. However, their relative contribution to interaction between charged biological helices, such as DNA, is not presently clear. The published theoretical studies of such interactions do not provide direct comparison with pertinent experimental results that are available in the literature. At the same time, predictions of the theory [27,30] that neglects surface charge density fluctuations but takes into account the discrete helical structure of surface charge distributions seem to be consistent with both repulsive and attractive interactions measured between double-stranded DNA helices and between four-stranded guanosine helices [27,30,35,36]. At the present, initial step of our study we will neglect forces associated with surface charge density fluctuations. However, the possibility that these forces may not be negligible is useful to keep in mind until this issue is clarified.

## A. Debye-Hückel-Bjerrum approximation

*Interaction in electrolyte solution.* Many biological macromolecules, DNA in particular, have high density of surface charges. Under physiological conditions, they are surrounded by an ionic atmosphere with the Debye length $\lambda_D \sim 7$ Å. It is natural to expect here a nonlinear screening of fixed surface charges by electrolyte ions which is often treated within the mean-field, nonlinear Poisson-Boltzmann theory [19,32,37]. However, most of counterions which contribute to the nonlinear screening lie within a narrow layer around each molecule. We refer to them as condensed counterions. The thickness of this layer can be estimated as $d_c \leq A/4\pi l_B$, where $A$ is the average area per elementary charge on the molecular surface and $l_B = e^2/\varepsilon k_B T$ is the Bjerrum length ($\approx 7$ Å in water). For most biological macromolecules $A \leq 100$ Å$^2$ and $d_c < 2$ Å. The thickness of this layer is comparable or even smaller than the size of a water molecule and than the characteristic roughness of the corrugated molecule/water interface. Mean-field description of an electric field inside such a thin layer is inappropriate.

We replace the Poisson-Boltzmann approximation by explicit treatment of condensed counterions. Specifically, we describe the molecular-core/water interface and the nonlinear screening layer as a single, infinitesimally thin surface that may have an arbitrary, inhomogeneous charge density. This surface contains fixed surface charges, chemisorbed ions, and mobile, condensed counterions. We describe the diffuse ionic atmosphere outside this surface within the Debye-Hückel theory [38].

This approach is similar to the two-state model (condensed counterions and free ions) commonly used in polyelectrolyte theory at nonvanishing salt concentrations [2,19,34]. It is also similar to the *Debye-Hückel-Bjerrum* model which has proved to be quite successful in the theory of concentrated electrolyte solutions, including the theory of Coulomb criticality [39-42]. In our case, one may expect such model to be accurate as long as the ratio of $d_c$ to all other characteristic lengths in the system (the Debye length, the surface-to-surface distance between the molecules, the helical pitch, etc.) is small.

*Interaction in nonpolar medium.* While nucleic acid helices (DNA or RNA) typically reside in aqueous electrolyte environment, many protein helices are immersed into nonpolar media. These may be α-helical domains inside a large globular protein or, e.g., transmembrane α-helices. Interaction between helical macromolecules in a nonpolar environment is relevant not only because of its potential biological applications, but also because of its conceptual importance. Indeed, lipid membranes and the interior of globular proteins have about the same dielectric constant as cores of α-helices, i.e. we can assume that the dielectric constant is the same everywhere. Furthermore, nonpolar media do not have dissolved electrolyte ions inside. This allows us to avoid approximations associated with interfaces between two different dielectrics and difficulties inherent to theories of electrolyte solutions. As a result, we can obtain **the exact** solution of the electrostatic problem as a particular case of the Debye-Hückel-Bjerrum model at $\kappa_D = 1/\lambda_D = 0$ and $\varepsilon_c = \varepsilon_s = \varepsilon$, where $\varepsilon_c$ is the dielectric constant of the molecular cores and $\varepsilon_s$ is the dielectric constant of the solvent.

## B. Surface charge patterns

We describe all charges at each molecular-core/water interface explicitly by their surface charge densities $\sigma_\nu(z,\phi)$, each in its own "molecular" frame of cylindrical coordinates. The index $\nu(=1,2)$ labels the two molecules. The $z$ axis of each molecular frame coincides with the molecular axis; $z=0$ is the point of the closest approach between molecular axes; $\phi=0$ corresponds to the direction of the vector **R** connecting the points of the closest approach on the axes of the two molecules (Fig. 1). In the Appendix, we derive a general relationship between the interaction energy and the cylindrical Fourier transforms of the surface charge densities $\tilde{\sigma}_\nu(q,n)$ that are defined as follows

$$\tilde{\sigma}_\nu(q,n) = \frac{1}{2\pi} \int_0^{2\pi} d\phi \int_{-\infty}^{\infty} dz\, \sigma_\nu(z,\phi)\, e^{in\phi} e^{iqz}, \qquad (1)$$

In the present work we focus on interaction between two identical helical macromolecules assuming that their surface charge densities obey the basic helical symmetry requirement

$$\sigma_\nu(z+\Delta z, \phi + g\Delta z) = \sigma_\nu(z,\phi), \qquad (2)$$

where $g=2\pi/H$ for right-handed helices, $g=-2\pi/H$ for left-handed helices, and $H$ is the helical pitch. Typically, this occurs when the charge density follows the underlying symmetry of the molecule. The surface charge density can, then, be rewritten in the form

$$\sigma_\nu(z,\phi) = \sigma_0 P(z - \phi/g - z_\nu + \phi_\nu/g). \qquad (3)$$

Here $P(x)$ is a periodic function that defines the axial charge distribution,

$$P(x + Hk) = P(x); \qquad (4)$$

$z_\nu$ and $\phi_\nu$ define relative dispositions of the molecules; $\sigma_0$ is the mean surface density of dominant [43] fixed charges. For example, we describe a DNA molecule by the coordinate $z_\nu$ of the center point on its principal axis and by the azimuthal angle $\phi_\nu$ of the middle of the minor groove at $z=z_\nu$ (Fig. 2).

Within this definition, the mean surface charge density is

$$\bar{\sigma} = \sigma_0 \frac{1}{H} \int_0^H P(x)\,dx. \qquad (5)$$

The cylindrical Fourier transform of $\sigma_\nu(z,\phi)$, defined by Eq. (3), gives

$$\tilde{\sigma}_\nu(q,n) = 2\pi\sigma_0 e^{ingz_\nu - in\phi_\nu} p(q)\delta(q+ng), \qquad (6)$$

where $\delta(x)$ is the Dirac's delta-function and

$$p(q) = \frac{1}{H} \int_0^H dx\, e^{iqx}\, P(x) \tag{7}$$

The reference point coordinates $(z_v, \phi_v)$ can always be selected to ensure real values of $p(q)$ at all $q$. For DNA, this is done by selecting the coordinate of the middle of the minor groove as $\phi_v$.

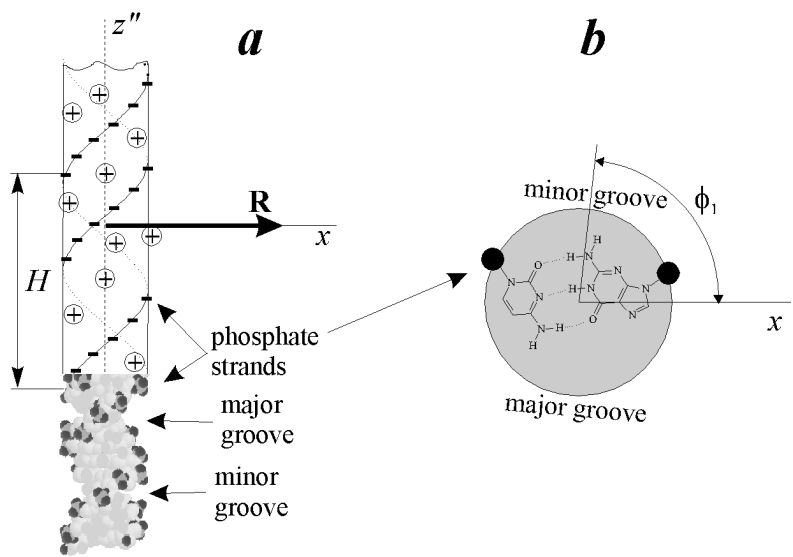

Fig. 2. (*a*) Structure of *B*-DNA with negatively charged oxygens of phosphate groups shown in black (bottom) and the corresponding model of charge distribution on DNA surface (top). Adsorbed (condensed) counterions are shown schematically as residing in the major and minor groves. The model represents molecule 1 and it is shown in the projection on $(x'',z'')$ plane of the corresponding molecular frame (c.f. Fig. 1). **R** is the vector connecting the points of the closest approach on the axes of molecules 1 and 2 (Fig. 1) where molecule 2 is not shown. (*b*) Cross-section of the molecule at $z''=z_1$. The azimuthal orientation of the molecule is defined by the azimuthal coordinate of the middle of the minor groove $\phi_1$.

## C. *B*-DNA helices

The *B*-form (see Fig. 2) is the most common state of DNA in aqueous solutions [44]. *B*-DNA has an inner core formed by hydrogen bonded Watson-Crick nucleotide pairs. The nucleotides are attached to two sugar-phosphate strands spiraling around the core and forming the well-known double helix that has the $H \approx 33.8$ Å pitch ($g \approx 0.186$ Å$^{-1}$). Each phosphate group bears a negative charge. Their centers lie at the radial distance $a \approx 9$ Å from the molecular axis. There are two phosphate groups per base pair and $\approx 10$ base pairs per helical turn so that $\sigma_0 \approx 16.8$ $\mu$C/cm$^2$. The helix is fairly stiff, its persistence length is 500-1000 Å (depending on the ionic strength) [see, e.g., 45 and refs. therein]. In cells and viruses, DNA is often packed within a tight compartment so that different folds of the same helix or different helices come to ~5-30 Å surface-to-surface separation within each other ($R$~25-50 Å) and form various liquid crystalline phases [4,5].

Since simple structural and dimensional arguments did not explain chiral properties of liquid crystalline phases of DNA, a natural approach is to develop a statistical theory based on a chiral pair potential. Here we calculate such potential based on the simplest possible model of DNA surface charge pattern that incorporates molecular chirality. Specifically, we consider phosphate strands as continuous, charged helical lines. We distinguish four qualitatively different types of location of adsorbed and condensed counterions: (1) on a helical line in the middle of the minor groove, (2) on a helical line in the middle of the major groove, (3) on phosphate strands, (4) random. From the corresponding model for $P(z)$, we find [29,30]

$$p(q) = (1 - f_1 - f_2 - f_3)\theta\delta_{q,0} + f_1\theta + f_2\theta\cos(\pi q/g) - (1 - f_3\theta)\cos(\tilde{\phi}_s q/g) \quad (8)$$

where $\delta_{x,y}$ is the Kroenecker's delta; $\tilde{\phi}_s$ ($\approx 0.4\pi$) is the azimuthal half-width of the minor groove; $\theta$ is the fraction of the charge of phosphates neutralized by adsorbed and condensed counterions; $f_i$ are the fractions of counterions in the middle of the minor groove ($f_1$), in the middle of the major groove ($f_2$), and on the phosphate strands ($f_3$); $f_1 + f_2 + f_3 \leq 1$. The fraction of randomly adsorbed counterions is $(1 - f_1 - f_2 - f_3)$. As a reference point for $P(z)$ we select the middle of the minor groove (Fig. 2).

Of course, this model is an idealization but a reasonable one, at least for 150 bp DNA fragments. Indeed: (i) Such fragments are about one persistence length long (500 Å) so that they can be roughly considered as straight and rigid. (ii) As long as the energy of interaction between molecules exceeds the thermal energy ($k_BT$), thermal motions may introduce corrections, but they should not change the overall qualitative behavior of the liquid crystalline phase. (iii) For 500 Å long fragments, base pair heterogeneity does not lead to significant deviations from the ideal helix [46]. (iv) For monovalent and polymeric counterions (such as polyamines) that were used to study the cholesteric phase, discreteness of phosphate and counterion charges is not essential, e.g., the helical chain of discrete phosphates can be modeled as a continuous charged spiral [47]. (v) This model has already proved to be quite successful in several applications [29,30].

### III. GENERAL RESULTS

#### A. Interaction at nonzero interaxial angles

*In an electrolyte solution*, the energy of interaction between two identical, infinitely long helices crossing at an interaxial angle $\psi \neq 0$ is given by (for derivation see Appendix)

$$E_{int} \approx \frac{8\pi^3 \sigma_0^2}{\varepsilon_s |\sin\psi|} \sum_{n,m=-\infty}^{\infty} \frac{(-1)^m p(-ng) p(mg) \cos[n(\phi_1 - gz_1) - m(\phi_2 - gz_2)] e^{-\kappa_m R\sqrt{1+w_{n,m}^2(\psi)}}}{\kappa_n \kappa_m^2 \sqrt{1+w_{n,m}^2(\psi)} (1-\tilde{\beta}_n(ng))(1-\tilde{\beta}_m(mg)) K'_n(\kappa_n a) K'_m(\kappa_m a)} \quad , \quad (9)$$

$$\times \left(\sqrt{1+w_{m,n}^2(\psi)} + w_{m,n}(\psi)\right)^n \left(\sqrt{1+w_{n,m}^2(\psi)} + w_{n,m}(\psi)\right)^m$$

where

$$w_{n,m}(\psi) = \frac{ng - mg\cos\psi}{\kappa_m \sin\psi}; \qquad (10)$$

$$\kappa_n = \sqrt{\kappa^2 + n^2 g^2}; \qquad (11)$$

$$\tilde{\beta}_n(q) = \frac{\varepsilon_c}{\varepsilon_s} \frac{|q|}{\kappa_n} \frac{K_n(\kappa_n a) I'_n(|q|a)}{I_n(|q|a) K'_n(\kappa_n a)}; \qquad (12)$$

$a$ is the radius of the water-impermeable molecular cores of the helices; $R$ is the closest approach distance between molecular axes (Fig. 1); $I_m(x)$ and $K_m(x)$, $I'_m(x)$ and $K'_m(x)$ are the modified Bessel functions and their derivatives, respectively. The derivation of Eqs. (9)-(12) involves only one approximation, namely the truncation of the series of consecutive images after the first-order term (see Appendix). Relative contributions of higher-order images decrease as $\exp(-2(R-2a)/\lambda_D)$, where $R-2a$ is the closest approach surface separation between helices. At $R-2a > \lambda_D$, such approximation is sufficiently accurate for all practical purposes.

In further analysis of interaction between biological helices in an electrolyte solution we use that $\varepsilon_c \approx 2$, $\varepsilon_s \approx 80$, and $\tilde{\beta}_n \sim \varepsilon_c/\varepsilon_s << 1$. Therefore, we neglect neglect $\tilde{\beta}_n$ compared to 1. However, in a nonpolar medium $\tilde{\beta}_n \sim 1$ and it cannot be neglected.

*In a nonpolar medium*, after substitution of $\varepsilon_c = \varepsilon_s = \varepsilon$ and $\kappa = 0$ into Eq. (9), we find that the energy of interaction between crossed helices is given by

$$E_{int} = \frac{8\pi^3 \sigma_0^2 a^2}{\varepsilon |\sin\psi|} \sum_{n,m=-\infty}^{\infty} \frac{(-1)^m p(-ng) p(mg) \cos\left[n(\phi_1 - gz_1) - m(\phi_2 - gz_2)\right]}{|mg|} I_n(|nga|) I_m(|mga|)$$

$$\times \frac{\left(\sqrt{1+\tilde{w}_{m,n}^2(\psi)} + \tilde{w}_{m,n}(\psi)\right)^n \left(\sqrt{1+\tilde{w}_{n,m}^2(\psi)} + \tilde{w}_{n,m}(\psi)\right)^m e^{-|mg|R\sqrt{1+\tilde{w}_{n,m}^2(\psi)}}}{\sqrt{1+\tilde{w}_{n,m}^2(\psi)}} \qquad , (13)$$

where

$$\tilde{w}_{n,m}(\psi) = \frac{ng - mg\cos\psi}{|mg|\sin\psi}. \qquad (14)$$

This is *the exact* expression for the given choice of the helical surface charge pattern.

**B. Interaction between parallel helices**

At $\psi \to 0$, the energy of interaction between infinitely long molecules diverges because the molecules overlap over an infinite length. In this case, the meaningful value is the energy density per unit length which can be calculated as $\lim_{L\to\infty}(E_{int}/L)$, where $L$ is the length of the molecules. In other words, for infinitely long molecules, the $\psi=0$ and $\psi\neq 0$ cases should be treated separately. A general expression for the interaction energy at $\psi=0$ was derived in [27]. It is also recovered here in the Appendix (Eq. (A 52)). For helices with surface charge patterns given by Eq. (6), the energy density takes the form

$$\frac{E_{int}}{L} \approx \frac{8\pi^2\sigma_0^2}{\varepsilon_s} \sum_{n=-\infty}^{\infty} (-1)^n \frac{[p(ng)]^2 \cos[n(\Delta\phi - g\Delta z)] K_0(\kappa_n R)}{\kappa_n^2 [K'_n(\kappa_n a)]^2} , \qquad (15)$$

in an electrolyte solution and

$$\frac{E_{int}}{L} = \frac{8\pi^2\sigma_0^2 a^2}{\varepsilon} \sum_{n=-\infty}^{\infty} (-1)^n [p(ng)]^2 \cos[n(\Delta\phi - g\Delta z)] K_0(|ng|R) I_n^2(|ng|a) \qquad (16)$$

in a nonpolar medium. Here $\Delta z = z_2 - z_1$, $\Delta\phi = \phi_2 - \phi_1$.

## C. Interaction modes

Although they may seem to be cumbersome, Eqs. (9)-(16) are easy to use. Indeed, it follows from Eqs. (9),(13),(15),(16) that the total interaction energy can be represented as a sum of contributions from "interaction modes" with different indices $n$ and $m$. Only a few modes with small indices give significant contributions to the total energy. The sum of modes rapidly converges because of the exponential dependence of the energy of each mode on $n$ and $m$. Unless $p(ng)$ is zero or anomalously small because of a peculiar symmetry of the charge pattern, the sum can be truncated after only a few terms with $n,m\neq 0$. The truncated expressions allow us to conduct fairly detailed analysis of the interaction laws, as well as to perform rapid numerical calculation.

For instance, for homogeneously charged cylinders, $\bar\sigma = \sigma_0$ and $p(q) = \delta_{q,0}$. After substituting this into Eq. (9) we arrive at the classical result [17,18,48]

$$E_{int} \approx \frac{8\pi^3\bar\sigma^2}{\varepsilon_s \kappa^3 K_1^2(\kappa a)} \frac{e^{-\kappa R}}{|\sin\psi|} . \qquad (17)$$

However, such approximation may not always be appropriate for helical macromolecules. It accounts for just a fraction of the net interaction energy. Most importantly, it neglects the chiral nature of helices and, therefore, chiral interactions.

For helices, the zero mode is nothing else as the contribution from the average surface charge density $\bar\sigma$. Chiral interactions are determined by other, "helical" modes with $n,m\neq 0$. Not only the helical modes are responsible for the chirality, but they often give a dominant contribution to the net energy. In particular, net-neutral molecules contain equal number of negatively and positively charged surface residues so that $\bar\sigma = 0$ while $\sigma_0 \neq 0$. Then, the interaction energy is determined exclusively by the helical modes. Molecules with high surface density of intrinsic charged residues of one sign (such as DNA) cause adsorption and/or condensation of counterions resulting in $|\bar{\sigma}|<<|\sigma_0|$. As a result, "helical" modes may contribute much more to the interaction energy than the zero mode (at surface separations smaller than the helical pitch).

## IV. INTERACTION BETWEEN INFINITELY LONG HELICES

### A. Interaction between idealized *B*-DNA-like helices

First, let us illustrate the general formulae with a heuristic model of interaction between *B*-DNA-like helices. We exaggerate the rigidity, homogeneity, and length of DNA and assume that the molecules are perfect, infinitely long helices. To emphasize nontrivial features of their chiral interaction, we select a peculiar (but not impossible) surface charge pattern, i.e. $f_1=0.4$, $f_2=0.6$, $f_3=0$, and $\theta=0.8$ in Eq. (8). In Section VI, we discuss observed phenomena using a model of 150 bp DNA fragments with surface charge patterns expected to be more common.

Several energy landscapes calculated from Eq. (9) within these assumptions are shown in Fig. 3. The energy is a function of six variables: interaxial separation $R$, interaxial angle $\psi$, angles of rotation about molecular axis $\phi_1$ and $\phi_2$, and axial shifts $z_1$ and $z_2$. It is plotted at two different surface separations: one slightly larger than the Debye length ($R=27$ Å, $\lambda_D=7$ Å, $R-2a\approx 9$ Å, [49]) and the other slightly larger than $2\lambda_D$ ($R=35$ Å). Each landscape shows the energy as a function of $\psi$ and $\Delta\phi=\phi_2-\phi_1$ at $z_1=z_2=0$ (i.e. when two very long molecules cross in the middle). The selected values of $\phi_1$ correspond to three most representative cases that differ in the location of the point of closest approach between the molecules with respect to strands and grooves on molecule 1. Specifically the point of the closest approach on molecule 1 is located in the center of the minor grove at $\phi_1=0$, on one of the phosphate strands at $\phi_1\approx 0.4\pi$, and in the center of the major groove at $\phi_1=\pi$ (see Fig. 2).

Numerical summation of the terms in Eq. (9) confirms our expectation of rapid convergence of the sums. We find that only the modes with $n,m=0,\pm 1,\pm 2$ are important. The modes of interaction with larger indices introduce minor corrections to the energy and they can be neglected.

The energy landscapes contain deep "canyons", shallow "lakes", "ridges", "overpasses", and steep "mountains". The interaction may be energetically favorable ($E_{int}<0$) or unfavorable ($E_{int}>0$) depending on the specific alignment of the molecules and on the interaxial separation. The point of plotting all these landscapes is to demonstrate the rich variety of their features. This richness suggests that the interaction may not be reducible to simple Hamiltonians used in phenomenological models of chiral interactions.

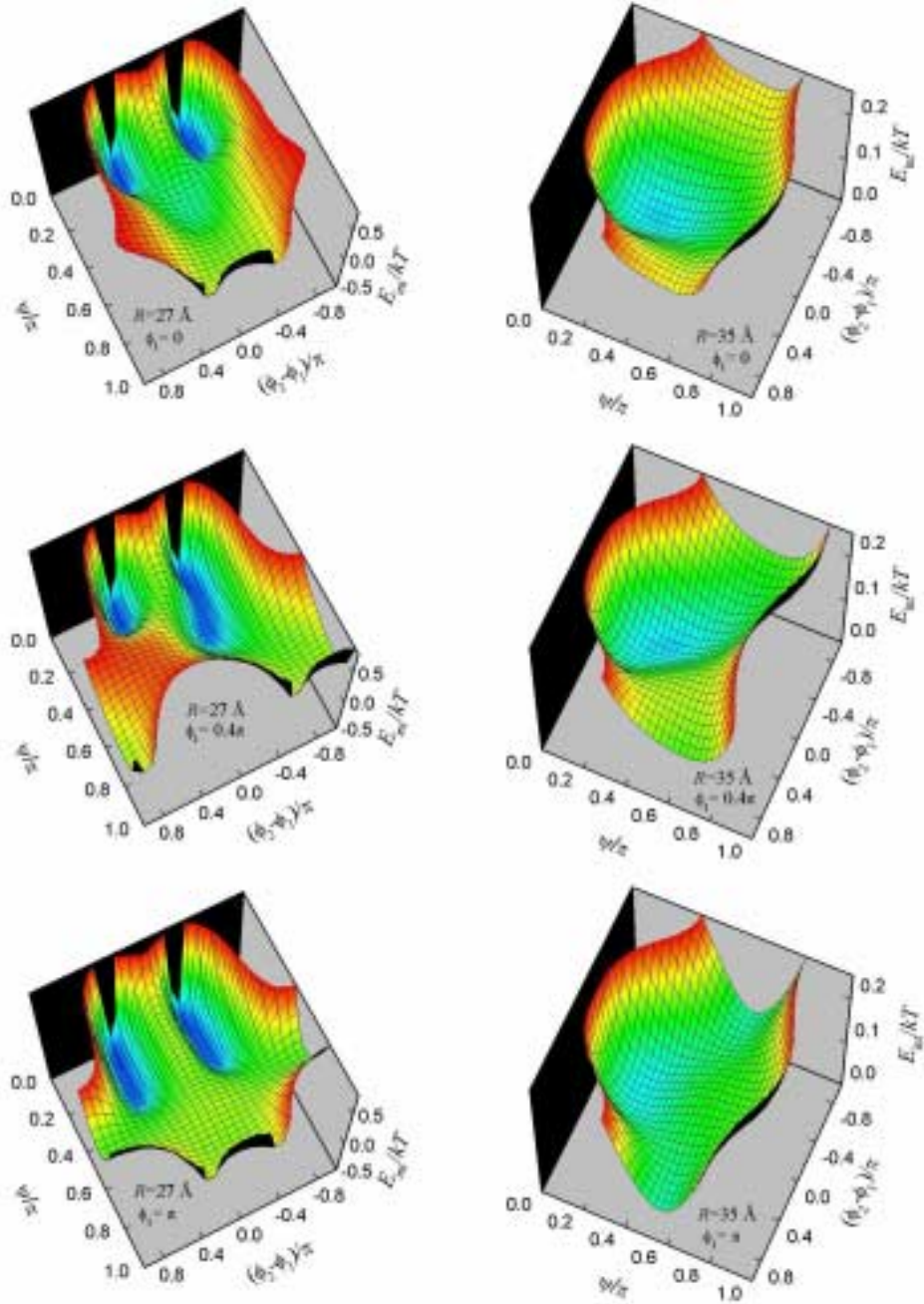

Fig. 3. Energy landscapes for interaction between two crossed *B*-DNA-like helices plotted as a function of the interaxial angle $\psi$ and the difference in about-axis-rotation angles $\phi_2-\phi_1$ for indicated values of $\phi_1$ and *R*. The energy is given in the units of thermal energy ($k_BT \approx 4.1 \cdot 10^{-14}$ erg $\approx 0.025$ eV at *T*=300 K). For infinitely long molecules it diverges as $\sim 1/|\sin\psi|$ at small interaxial angles because of the increasing effective interaction length (Section V.A). For finite-length molecules, the divergence levels off. Here, it is cut off from the landscapes by limiting the energy scale. To improve visual perception, the landscapes are shown for $\psi$ varying from 0 to $\pi$. Because of molecular symmetry, the ($\phi_1,-\psi$) alignment of molecule 1 is equivalent to ($2\pi-\phi_1,\pi-\psi$), e.g., $-\psi$ is equivalent to $\pi-\psi$ at $\phi_1=0$ and at $\phi_1=\pi$.

Revised 4/21/00    page 12

## B. Interaction at large separations

Details of the energy landscapes shown in Fig. 3 are model-specific, but the interaction also has universal features common for all helices. To gain an insight into such features, it is instructive to consider the interaction in a simple limiting case of small interaxial angles ($|\sin\psi|<<1$) and large surface separations ($R-2a>>\lambda_D$). At $|\sin\psi|<<1$, only $n=m$ terms contribute to the sum in Eq. (9) since $w_{n,m}(\psi)|_{n\neq m}>>1$ and $w_{n,n}(\psi)<<1$. When, in addition, $\kappa(R-2a)>>1$ all modes except those with $n=m=0$ and $n=m=\pm 1$ can be neglected [50]. After expanding the energy at small $|\psi|$, we find

$$\frac{\varepsilon_s \kappa^3 K_1^2(\kappa a)}{8\pi^3 \sigma_0^2} E_{int} \approx \left[\frac{\bar{\sigma}^2}{\sigma_0^2} - C\cos(\Delta\phi - g\Delta z)e^{-(\kappa_1-\kappa)R}\right] e^{-\kappa R}\frac{1}{|\psi|}$$
$$- \frac{Cg}{\kappa_1}\cos(g\Delta z)e^{-\kappa_1 R}\frac{\psi}{|\psi|} + \frac{Cg^2}{8\kappa_1^2}\cos(\Delta\phi - g\Delta z)(\kappa_1 R - 3)e^{-\kappa_1 R}|\psi| + ... \quad (18)$$

where $\bar{\sigma}$ is the average surface charge density [see Eq. (5)] and

$$C = 2[p(g)]^2 \frac{\kappa^3 K_1^2(\kappa a)}{\kappa_1^3 [K_1'(\kappa_1 a)]^2} \quad . \quad (19)$$

It follows from Eq. (18) that the most energetically favorable axial alignment between the molecules is $\Delta\phi=g\Delta z$, regardless of the interaxial angle. At this optimal alignment, the molecular opposition may be energetically favorable ($E_{int}<0$) or unfavorable ($E_{int}>0$), depending on the ratio $\bar{\sigma}/\sigma_0$ and on the surface separation between molecules.

At $|\bar{\sigma}|>|\sigma_c|$, where [51]

$$\sigma_c = \sigma_0 \sqrt{C}\, e^{-(\kappa_1-\kappa)a}, \quad (20)$$

the interaction is energetically unfavorable at zero and any small angle. The energy decreases with increasing interaxial angle.

When $|\bar{\sigma}|<|\sigma_c|$, the interaction is more complicated. It is energetically favorable at $R<R_c$, where

$$R_c = 2a + (\kappa_1 - \kappa)^{-1}\ln\left[\frac{\sigma_c^2}{\bar{\sigma}^2}\right]. \quad (21)$$

Within this distance range, the most favorable interaxial angle is $\psi=0$. The molecules "recognize" each other and tend to aggregate in a conformation with parallel long axes even when both molecules have non-zero mean surface charge density. This increases the effective length of the energetically favorable molecular opposition. The attraction is due to axial charge separation. The latter allows such alignment of two helices that oppositely charged surface groups face each other, as discussed previously in the theory of interaction between parallel helices [27,30]. At $|\bar{\sigma}|\to 0$, $R_c\to\infty$.

At $R>R_c$ but $R-R_c\ll R_c$, the interaction is energetically unfavorable at parallel alignment ($\psi=0$), but a small twist makes it favorable. The optimal angle is

$$\psi \approx \frac{2\kappa_1}{g}\sqrt{\frac{2\left[e^{(\kappa_1-\kappa)(R-R_c)}-1\right]}{(\kappa_1 R - 3)}} \qquad (22)$$

It increases with increasing separation. At larger $R$, the optimal interaxial angle goes outside of the range of the small angle approximation. The dependence of the optimal angle on $R$ in the vicinity of $R_c$ is shown in Fig. 4.

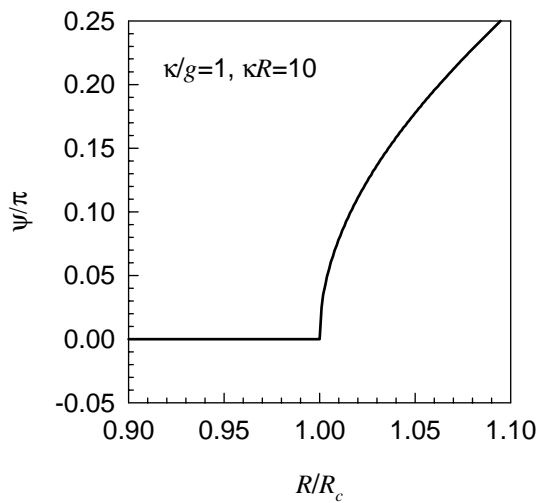

Fig. 4. The dependence of the most energetically favorable interaxial angle on interaxial separation in the asymptotic limit of large separations and small angles. The nonzero interaxial angle emerges continuously after a critical distance $R_c$ (c.f. Eq. (22)). The value of $R_c$ depends on the ratio between positively and negatively charges on molecular surface as described by Eq. (21). The stronger is the charge compensation, the larger is $R_c$.

Thus, helices tend to align parallel when they attract each other. They tend to cross at $\psi\sim\pi/2$ when they repel each other. This is something one would expect for any long, rod-like molecules. Such behavior was also predicted for homogeneously charged rods whose attraction was due to correlated surface charge density fluctuations [19]. The nontrivial conclusions of the present study are the following: (*a*) As a result of helical surface charge pattern, two molecules can attract each other and aggregate in parallel conformation even when they have nonzero net charges of the same sign. We first described this in [27,30]. (*b*) The transition between $\psi=0$ and $\psi\sim\pi/2$ upon the loss of attraction occurs gradually rather than as a jump (Fig. 4). A small twist extends the range of intermolecular attraction. (*c*) At small interaxial angles, chirality of helices affects the direction *but not the amplitude* of the most favorable twist. Indeed, the only term in Eq. (18) that depends on the sign of $\psi$ does not depend on the amplitude of $\psi$.

## V. INTERACTION BETWEEN HELICES OF FINITE LENGTH

There are no infinitely long helices in real life. Below we use a simple approximation of long ($L/H\gg 1$) helices to extend the theory to the case of interaction between molecules of finite length. We consider only a straightforward extension that can be done at ψ=0 and at Lψ larger than other relevant length scales, e.g., $R$, $\lambda_D$, $H$. Despite its limitations, this approach allows us to arrive at a number of interesting conclusions. A more general theory for all interaxial angles is a hard mathematical problem which is currently under investigation.

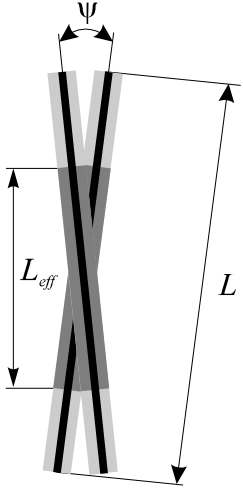

Fig. 5. Schematic illustration of interaction between two crossed, long molecules of finite length $L$ in projection on the $(y,z)$ plane (c.f. Fig. 1). Here the molecules are represented by solid black lines while light gray shading symbolizes electric fields around them. The region shaded in dark gray symbolizes energetically significant overlap between the electric fields. Since each $n$-th harmonic of the charge distribution on helical surface produces an electric field that decays exponentially away from the helix with the charecteristic decay length $\kappa_n^{-1}$, the length of the overlap region for each mode is $\sim \kappa_n^{-1}/|\sin\psi|$. This is the "effective interaction length" for the mode. Only a few modes contribute significantly to the interaction. The tips of helices do not contribute much to the interaction when they protrude beyond the overlap region for all essential modes (as shown). Then, the interaction energy can be calculated assuming that the helices are infinitely long. In the opposite limit of much smaller angles, the helices are effectively parallel.

## A. Effective interaction length

Consider a simple argument schematically illustrated in Fig. 5. Each $n$-th harmonic of the surface charge density produces an electric field that decays away from the helix with the characteristic length $\kappa_n^{-1}$. The fields created by two crossed infinitely long helices overlap over the length $L_n(\psi) \sim \kappa_n^{-1}/|\sin\psi|$, as it follows immediately from the geometry of the crossed configuration. At $|\sin\psi|\ll 1$, the helices are essentially parallel in the region of their overlap. Therefore, we should be able to represent the energy of their interaction as

$$E_{\text{int}} = \sum_{n=-\infty}^{\infty} L_n(\psi) u_n(\psi), \qquad (23)$$

where $u_n(\psi=0)$ is the energy of interaction between parallel helices per unit of their length.

Comparing Eq. (9) with Eq. (15) we find that the expression for $E_{\text{int}}$ at small ψ can be rewritten in the form of Eq. (23), where

$$u_n(\psi) \approx (-1)^n \frac{8\pi^2\sigma_0^2}{\varepsilon_s} \frac{[p(ng)]^2 \cos[n(\Delta\phi - g\Delta z)] K_0(\kappa_n R)}{\kappa_n^2 [K_n'(\kappa_n a)]^2}$$
$$\times \left[1 + n^2 \frac{g\sin\psi}{\kappa_n} - (\kappa_n R + 1 - 4n^2)\frac{n^2 g^2 \sin^2\psi}{8\kappa_n^2} + ...\right] \qquad (24)$$

and the *effective interaction length* for each mode $n$ is given by

$$L_n(\psi \neq 0) = \frac{\kappa_n^{-1}}{|\sin\psi|} \frac{\pi e^{-\kappa_n R}}{K_0(\kappa_n R)}. \tag{25}$$

These expressions apply also to helices of finite length if molecular tips protrude beyond the area of energetically significant overlap between the electric fields, as shown in Fig. 5. Then the contribution of the tips to the interaction energy can be neglected. Since the length of the overlap area for each mode is $L_n(\psi)$, this approximation should work when $L>L_n(\psi)$ for all essential modes. Taking into account that $L_0(\psi)$ is the largest of all $L_n$, we find that Eqs. (23)-(25) are applicable to helices of finite length at $|\psi|>\psi_0$, where $\psi_0 \approx \sqrt{2\pi R/\kappa}/L$ is the root of equation $L_n(\psi)=L$. To obtain this expression from Eq. (25), we replaced $|\sin\psi|$ by $|\psi|$ and used the asymptotic expansion of $K_0(x)$ assuming that $\kappa R>>1$, which is typically the case.

Furthermore, Eqs. (23),(24) also describe the energy of interaction between parallel ($\psi=0$) helices of large finite length $L$ ($L>>H$) if we adopt

$$L_n(\psi=0)=L. \tag{26}$$

Indeed, two helices can be viewed as effectively parallel over their whole length when $L<<L_n(\psi)$ for all essential modes.

Thus, we can use Eqs. (23)-(26) for finite length helices as long as $L>>H$ and $|\psi|<<\psi_N$ or $|\psi|>\psi_0$. Here $N$ is the largest mode index that still contributes significantly to the interaction energy (e.g., $N=2$ for interaction between *B*-DNA helices, see Fig. 7) and

$$\psi_n \approx \frac{\pi}{L\kappa_n} \frac{e^{-\kappa_n R}}{K_0(\kappa_n R)} \approx \frac{\sqrt{2\pi R/\kappa_n}}{L} \tag{27}$$

The intermediate case of $|\psi|\sim\psi_n$ is more complicated. Its rigorous analysis requires a full, explicit solution for the energy of interaction between crossed helices of finite length at all interaxial angles. This is still an unsolved mathematical problem currently under consideration.

## B. Intermolecular torque

It was proposed that to understand the macroscopic pitch of a cholesteric phase formed by long, chiral molecules it is sufficient to know the intermolecular torque at $\psi=0$ [14,15]. This rests on the assumption that the torque is a regular, smooth function of $\psi$ at small $\psi$. Let us now calculate the torque between two helices and see whether our results support such assumption.

At $\psi=0$, we find from Eqs. (24),(26) that the intermolecular torque is

$$t = -\frac{dE_{\text{int}}}{d\psi} \approx -\frac{16\pi^2\sigma_0^2}{\varepsilon_s} Lg \sum_{n=1}^{\infty}(-1)^n n^2 \frac{[p(ng)]^2 \cos[n(\Delta\phi - g\Delta z)] K_0(\kappa_n R)}{\kappa_n^3 [K_n'(\kappa_n a)]^2}. \tag{28}$$

As we would expect, the torque changes the sign upon inversion of helical handedness ($g>0$ for right-handed and $g<0$ for left-handed helices). This torque tends to twist helices out of parallel alignment.

At $\psi_0 < |\psi| \ll 1$, we obtain from Eqs. (24),(25)

$$t \approx \frac{\psi}{|\psi|^3} \frac{8\pi^3 \sigma_0^2}{\varepsilon_s} \sum_{n=-\infty}^{\infty} (-1)^n \frac{[p(ng)]^2 \cos[n(\Delta\phi - g\Delta z)] e^{-\kappa_n R}}{\kappa_n^3 [K'_n(\kappa_n a)]^2} \tag{29}$$

An extrapolation of Eq. (29) to $\psi=0$ yields a diverging torque instead of Eq. (28). We illustrate possible consequences of such behavior below on the example of the chiral phase formed by 150 base-pair-long DNA helices.

## VI. ON THE ORIGIN AND STRUCTURE OF THE CHOLESTERIC PHASE FORMED BY *B*-DNA MOLECULES

As we discussed in Section II.C, the model of *B*-DNA as a rigid rod with helical surface charge distribution (Eq. (8)) should be reasonably accurate for describing electrostatic interactions between molecules whose total length is large compared to the helical pitch but smaller or equal to the persistence length. This applies to 150 base pair fragments of *B*-DNA molecules whose length is ~ 500 Å. The phase behavior of concentrated solutions of such fragments was extensively studied experimentally [2,3,5,6, 52, 53]. Most of the studies were performed in ~0.1-0.3 M NaCl or $NH_4Cl$ ($\lambda_D$~5-10 Å, $\theta \approx 0.75$ [54]).

The observed sequence of phases upon increasing water content (increasing interaxial separation, $R$) is: (1) a non-chiral, hexagonal [55] phase at $R<32$ Å, (2) a chiral, cholesteric phase at $R$ from 32 Å to 49 Å, and (3) an isotropic phase at $R> 49$ Å [2,52]. The averaged structure of the cholesteric phase can be represented by a helically twisted stack of layers, each formed by parallel molecules (Fig. 6). The twist angle between adjacent layers is 0.3 deg to 3 deg producing a very large cholesteric pitch $P$~0.4-5 $\mu$m [5,6,10]. Because of the small twist angle, the packing of nearest neighbor molecules in the cholesteric phase is almost hexagonal.

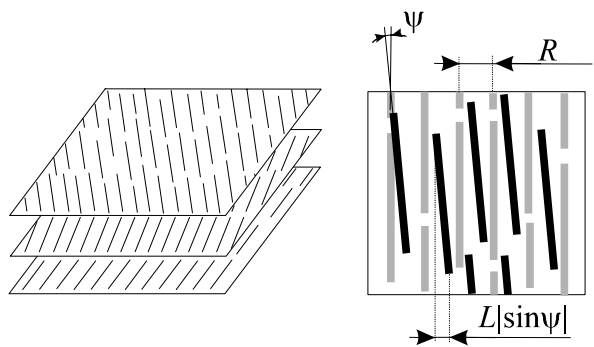

Fig. 6. *Left*. Schematic illustration of a right-handed cholesteric phase. *Right*. Mutual alignment of molecules in two neighboring layers in the cholesteric phase. Molecules in the bottom layer are shown in gray while molecules in the top layer are shown in black.

Ideally, in order to build a molecular theory of the cholesteric phase, one should solve a statistical many body problem using potentials of intermolecular forces and taking into account molecular motions (rotation about the prinicpal axes, axial translation, fluctuations of the interaxial angle, etc. ). This is a hard, yet unsolved problem. However, we can try to get a glimpse at the underlying physics via a qualitative analysis of the contributing forces, torques, and motions, as described below.

**A. Thermal motions**

The cholesteric arrangement of molecules is produced by a chiral torque (see, e.g., [15]). From osmotic stress measurements, we know that intermolecular interactions in the cholesteric phase of DNA are dominated by electrostatic forces [57]. We, therefore, assume that the torque has electrostatic origin as well [58]. Eqs. (28) and (29) describe the electrostatic torque at a fixed mutual alignment of molecules, i.e. in the absence of any molecular motions. From these expressions, one can see that thermal rotations of molecules about their principle axes and translations along the axes may have a profound effect on the average torque. For instance, chiral components of the torque can be completely wiped out if such thermal motions of neighboring molecules are not correlated. Knowledge of the strength of the corresponding correlations is therefore critical for understanding the cholesteric phase.

It is generally accepted that cholesteric phases exhibit long-range orientational order of molecules and short-range positional order [59]. It is also often presumed that molecules can almost freely rotate about their principle axes [60]. However, there is no experimental evidence that the latter is true for DNA. Furthermore, theoretical estimates favor an opposite picture. From Eqs. (23)-(26), we find that the energetic cost of rotation ($\delta\phi$) of a molecule surrounded by six nearest neighbors is $\sim B(R)\cos(\delta\phi)$ where $B(R) = 48\pi^2 L \sigma_0^2 \kappa_1^{-2} [K_1'(\kappa_1 a)]^{-2} \varepsilon_s^{-1} K_0(\kappa_1 R)$. In the middle of the range of existence of the cholesteric phase ($R \approx 40$ Å), we find $B \sim 5\ k_B T$ for 150 bp long DNA. The cost of molecular rotation is quite high suggesting that correlations in the mutual alignment of nearest neighbors should be rather strong.

Similarly, the cost of an axial shift ($\delta z$) of a molecule at constant $\phi$ is $\sim B(R)\cos(g\delta z)$, where $B(R)$ is the same as above, i.e. this motion should also be strongly hindered. However, a screw motion ($\delta\phi = g\delta z$) with a relatively small axial amplitude ($\delta z \ll L$) can proceed at almost no energetic cost [61]. Such motion does not affect mutual alignment of charge patterns on neighbor molecules (except for small edge effects). Therefore, the corresponding thermal excitations have no effect on the torques calculated above. However, these motions do change physical positions of molecules relative to each other.

In addition to about-principal-axis rotations and axial translations, intermolecular interactions may be affected by fluctuations of the interaxial distance, e.g., due to DNA undulations [62]. Such fluctuations double the decay length of the electrostatic interaction [63]. As a result, they should increase the effective value of $B(R)$ and hinder thermal rotations and translations even further. However, they should not have a significant effect on the relative balance of torques because they do not change the ($\Delta\phi, \Delta z$) alignment [62].

Using the above estimates as a guide, we assume that torques responsible for the cholesteric arrangement of DNA can be calculated in the "ground state" approximation for the mutual alignment of charge patterns on nearest neighbor molecules. Hindered thermal rotations and axial translations that disrupt the alignment introduce only minor corrections, except near the edge of the existence of the phase. However, this relatively rigid alignment of charge patterns on nearest neighbor molecules coexists with virtually unhindered, independent screw motions of each molecule and, therefore, with the absence of any long-range positional order in the phase. In the absence of better data or a more detailed theory, this is a logical assumption that we use in further analysis of the balance of forces and torques.

**B. Pairwise additivity and many-body effects.**

As we mentioned in Section II, here we consider only "frozen" surface charge patterns postponing analysis of possible effects of surface charge density fluctuations till later studies. Then, taking into account that electrostatic potentials are additive, we can find the ensemble energy by adding up energies of interaction of all pairs of DNA molecules. Furthermore, because surface separation between any two molecules in the cholesteric phase exceeds the Debye length and because the interaction energy exponentially decreases with the separation we can consider only interactions between nearest neighbors.

Despite the additivity of pair interaction potentials, the dependence of the energy on mutual alignment of molecules may introduce many-body effects. Consider, for instance, interaction between three nearest neighbor molecules. The energy of interaction between each two of them $\nu,\mu$ (=1,2,3) depends on their mutual alignment $(\Delta\phi-g\Delta z)_{\nu,\mu}$ (see Eq. (24)). The many body effects are due to the relationship between $(\Delta\phi-g\Delta z)_{\nu,\mu}$ that has the form

$$(\Delta\phi-g\Delta z)_{1,3}=(\Delta\phi-g\Delta z)_{1,2}-(\Delta\phi-g\Delta z)_{2,3}. \tag{30}$$

The cholesteric twist between adjacent molecular layers would have lead to even more complex many-body effects if a molecule in one layer crossed with more than one molecule in the preceding layer. However, there are no such multiple crossings in the cholesteric phase of *B*-DNA. At typical twist angles $\psi \leq 3$ deg, a molecule traverses the distance $L|\sin\psi| < 25$ Å across the adjacent layer that is less than the interaxial distance ($R>32$ Å) in the layer (Fig. 6).

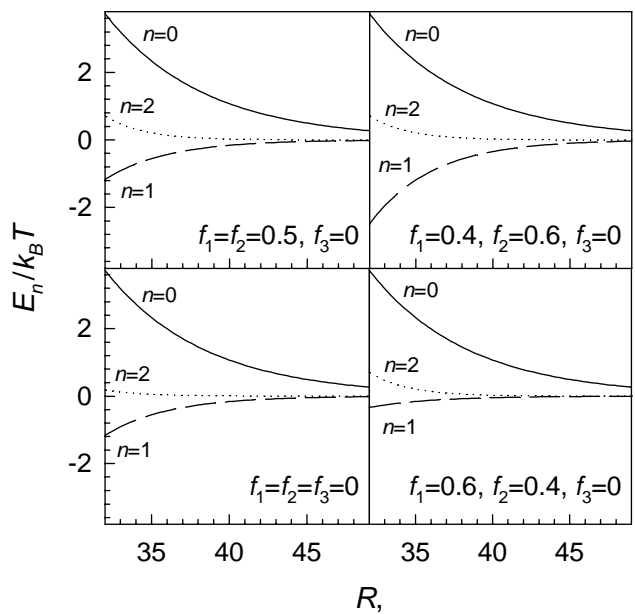

Fig. 7. Contribution of different modes to the energy of interaction between two parallel ($\psi$=0), 150 bp DNA helices at $\theta$=0.75. Each $E_n$ is the sum of the energies of two modes with indices $\pm n$. Relative contributions of different modes to the interaction energy at small $\psi \neq 0$ are practically the same as at $\psi$=0, as follows from Eq. (24). The energy of modes with $n=\pm 3$ is not shown since it is indistinguishable from zero at the energy scale used on the graph.

## C. Force and torque balance

Consider the balance of forces and torques in the cholesteric phase. First of all, note that only interaction modes with $n=m=0,\pm 1,\pm 2$ are important at $R\sim 32\text{-}49$ Å (see Fig. 7). Each mode has a characteristic interaxial angle $\psi_n$ which determines the upper boundary of the crossover region where intermolecular torque rapidly changes its behavior (Section V.B). From Eq. (27), we find that for B-DNA $\psi_0 \approx 5$ deg, $\psi_{\pm 1} \approx 4$ deg, and $\psi_{\pm 2} \approx 3$ deg. The angles observed in the cholesteric phase (0.3-3 deg), thus, lie exactly in the crossover region. Below we show that this may not be a coincidence.

At $|\psi| << \psi_2$, the molecules are effectively parallel and their local packing can be approximated by a hexagonal array. At such small angles, the energy of intermolecular interaction and the torque between molecular layers can be calculated in the limit of $\psi \to 0$. The interaction energy is determined primarily by the mode with $n=0$, but this mode is not chiral. From symmetry, nonchiral modes cannot produce a torque at $\psi=0$. Thus, the net torque in this range is chiral and it is determined by the modes with $n=\pm 1, \pm 2$. From Eqs. (8) and (28), we find the average interlayer torque per molecule,

$$\langle t \rangle \approx \frac{8\pi^2 \sqrt{6\pi} L g \sigma_0^2 e^{-\kappa_1 R}}{\varepsilon_s \kappa_1^3 \sqrt{\kappa_1 R}\left[K_1'(\kappa_1 a)\right]^2} \left\{ \left\langle \cos\left[(\Delta\phi - g\Delta z)_{\nu,\mu}\right]\right\rangle \left[(f_2 - f_1)\theta + \cos(\tilde{\phi}_s)\right]^2 \right.$$

$$\left. - 4\left\langle \cos\left[2(\Delta\phi - g\Delta z)_{\nu,\mu}\right]\right\rangle \left[(f_2 + f_1)\theta - \cos(2\tilde{\phi}_s)\right]^2 \frac{\kappa_1^{7/2}\left[K_1'(\kappa_1 a)\right]^2}{\kappa_2^{7/2}\left[K_2'(\kappa_2 a)\right]^2} e^{-(\kappa_2-\kappa_1)R} \right\}$$

(31)

where $< >$ indicates averaging over alignment between nearest neighbor molecules $\nu$ and $\mu$. In deriving Eq. (31), we assumed that $f_3=0$ (all counterions are located in the grooves) and used the asymptotic expansion of $K_0(\kappa_n R)$ at $\kappa_n R >>1$. This chiral torque is the driving force for formation of the cholesteric phase. It tends to establish either right-handed or left-handed twist between molecular layers depending on the average molecular alignment and on the partition of counterions between the grooves.

At $|\psi|>\psi_1$, tips of the molecules are separated by a sufficient distance so that the change in the effective interaction length (Fig. 5) becomes the main source of the energy change with $\psi$ (see Section V.A). As a result, the torque is determined by the mode that gives the dominant contribution to the energy, i.e. by the repulsion between unbalanced charges on molecular surfaces ($n=0$ mode). The intermolecular interaction becomes essentially the same as between homogeneously charged rods that tend to have parallel orientation and pack into hexagonal lattice because this increases average separation between molecular surfaces. The net torque becomes predominantly nonchiral and it favors a decrease in $|\psi|$, as described by Eq. (29).

The competition between the chiral torque (dominant at $|\psi|<<\psi_2$) and the nonchiral torque (dominant at $|\psi|>\psi_1$) produces a nonzero equilibrium interaxial angle $|\psi_{eq}|$ somewhere in the crossover region. To determine the exact value of $|\psi_{eq}|$, one needs to know the full angular dependence of intermolecular torque. As we noted in Section V.A, calculation of the interaction energy in the crossover region is still an unsolved mathematical problem. We, therefore, do not know the exact position of the optimum angle, the energy depth in the minimum, and the energy profile around the minimum. However, even without knowing the exact answer to these

questions, we can estimate the cholesteric pitch as follows.

## D. Cholesteric pitch

The cholesteric pitch ($P$) is related to $|\psi_{eq}|$ as $P=\sqrt{3}\pi R/|\psi_{eq}|$ [64]. Since $|\psi_{eq}|<\psi_1$, the pitch of the cholesteric phase should be larger than $\sqrt{3}\pi R_{min}/\psi_1$, where $R_{min}$ is the smallest interaxial separation between helices in the cholesteric phase. Since $|\psi_{eq}|$ is not much smaller than $\psi_2$, we assume that $|\psi_{eq}|$ and $\psi_2$ have the same order of magnitude or $|\psi_{eq}|>0.1\psi_2$. Therefore, we can expect $\sqrt{3}\pi R_{min}/\psi_1 < P < 10\sqrt{3}\pi R_{max}/\psi_2$ where $R_{max}$ is the largest interaxial separation between helices in the cholesteric phase.

Using $R_{min}\approx 32$ Å and $R_{max}\approx 49$ Å [52], $H\approx 34$ Å, $L\approx 500$ Å, and $\lambda_D\approx 5$-$10$ Å, we obtain the expected cholesteric pitch $0.3$ μm$<P<7$ μm. This is exactly the range observed in experiments. If the agreement is not a coincidence, it confirms that thermal rotations of molecules about their principal axes are not significant in the cholesteric phase of DNA. Such rotations would significantly reduce the strength of chiral interaction resulting in a much larger cholesteric pitch. At least at this level, our model appears to be self-consistent.

## E. Alignment frustration and cholesteric-to-hexagonal phase transition

It follows from Eq. (31) that the average chiral torque has a nonzero value only when molecular alignment is not random, i.e. $\left\langle \cos\left[ (\Delta\phi - g\Delta z)_{v,\mu} \right] \right\rangle \neq 0$. At random alignment the molecules are rotated about their principle axes by random angles so that the average molecular symmetry and average electrostatic interactions become effectively uniaxial [14,65]. The chiral cholesteric phase can not be formed as a result of uniaxial interactions that are always nonchiral [15,59]. Therefore, correlations in molecular alignment of helices play critical role in the cholesteric phase formation.

The ability of pairs of adjacent molecules to preserve their mutual alignment in a multimolecular environment depends on whether their optimal (most energetically favorable) pairwise alignment is compatible with the packing symmetry of their neighbors. Since packing of nearest neighbor DNA in the cholesteric phase is almost hexagonal, we find from Eq. (30) that only $(\Delta\phi-g\Delta z)=0$ and $(\Delta\phi-g\Delta z)=\pm 2\pi/3$ can be simultaneously optimal for all pairs of molecules.

The optimal alignment for a pair of opposing molecules can be determined by minimizing the energy given by Eqs. (23), (24) with respect to $(\Delta\phi-g\Delta z)$. Taking into account only the modes with $|n|\leq 2$, we find that $(\Delta\phi-g\Delta z)=\pm 2\pi/3$ is energetically unfavorable for any molecular pair while $(\Delta\phi-g\Delta z)=0$ is optimal only for pairs of molecules whose axes are separated by $R\geq R_*$, where

$$R_* \approx \frac{1}{\kappa_2 - \kappa_1} \ln \left\{ \frac{4\left[ (f_2+f_1)\theta - \cos(2\tilde{\phi}_s) \right]^2 \left[ K_1'(\kappa_1 a) \right]^2 \kappa_1^{5/2}}{\left[ (f_2-f_1)\theta + \cos(\tilde{\phi}_s) \right]^2 \left[ K_2'(\kappa_2 a) \right]^2 \kappa_2^{5/2}} \right\} \quad (32)$$

At $R<R_*$, the optimal value of $(\Delta\phi-g\Delta z)$ gradually increases with decreasing $R$.

Nonzero optimal $(\Delta\phi-g\Delta z)$ for molecular pairs would result in non-optimal average alignment of at least some pairs of nearest neighbor molecules in the cholesteric phase, causing an alignment frustration. The frustration may be resolved by an alignment that is not optimal for some or all pairs but that is still more energetically favorable than random alignment. If the energetic advantage of such structure is sufficient to compete with the entropic advantage of random alignment one may expect chiral interactions and the cholesteric phase to be preserved. If it is not, one may expect a transition to nonchiral hexagonal packing.

We initially derived an expression similar to Eq. (32) in Ref. [27] for two parallel double helices with randomly distributed condensed counterions. We pointed out that the value of $R_*$ is close to the observed transition point from cholesteric to hexagonal packing in DNA assemblies and speculated that this may not be a coincidence. This idea was further developed in Ref. 12 whose authors were first to suggest that the cholesteric-to-hexagonal transition at $R \approx 32$ Å and the absence of any twist in the hexagonal (line hexatic) phase may be caused by the alignment frustration. The validity of this interpretation still awaits its confirmation by a statistical theory of multimolecular aggregates based on the molecular interaction potential.

In the absence of such theory, we can only continue to speculate that the transition is expected to occur somewhere at $R \leq R_*$ when the pairwise alignment with $(\Delta\phi - g\Delta z)_{\nu,\mu} = 0$ becomes nonoptimal. The dependence of $R_*$ on the partitioning of counterions between DNA grooves is illustrated in Fig. 8a. For instance, at equal partitioning ($f_1 = f_2 = 0.5$) and $\lambda_D \approx 7$ Å, we find $R_* \approx 37$ Å. This is in good agreement with the experimental observation that the cholesteric-to-hexagonal transition occurs at $R \approx 32$ Å.

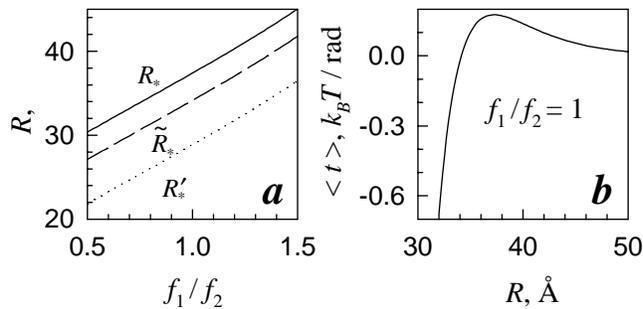

Fig. 8. Chiral torque between adjacent molecular layers in the cholesteric phase of 150 bp DNA at $|\psi| << \psi_2$ (3 deg) at fixed alignment of nearest neighbor molecules $(\Delta\phi - g\Delta z)_{\nu,\mu} = 0$. (*a*) Dependence of $R_*$, $R'_*$, and $\tilde{R}_*$ on the ratio of ions in the minor and major grooves ($f_1/f_2$). At $R < R_*$, the alignment with $\Delta\phi - g\Delta z = 0$ becomes non-optimal for individual pairs of molecules. At $R < R'_*$, the alignment with $\Delta\phi - g\Delta z = 0$ becomes energetically unfavorable compared to random alignment. At $R = \tilde{R}_*$, the average torque changes its sign. (*b*) Dependence of the average torque (normalized per one molecule) on interaxial distance at $f_1/f_2 = 1$.

### F. The sense of cholesteric twist

Intuitively, one expects a cholesteric phase formed by right-handed helices to have a right-handed cholesteric twist. The standard argument is illustrated in Fig. 9. In agreement with this intuitive expectation, Eq. (31) predicts a positive torque and, therefore, a right handed cholesteric twist at $R \geq R_*$ when the optimal alignment of all molecular pairs is $(\Delta\phi - g\Delta z) = 0$.

However, a counterintuitive torque behavior is possible at $R < R_*$. Consider, e.g., the behavior of the torque defined by Eq. (31) upon decreasing $R$ at fixed pairwise alignment $(\Delta\phi - g\Delta z) = 0$ (Fig. 8b). At first the torque is positive, but it changes its sign and becomes negative at $R < \tilde{R}_*$, where

$$\tilde{R}_* \approx R_* - \frac{\ln(\kappa_2/\kappa_1)}{\kappa_2 - \kappa_1} \ . \tag{33}$$

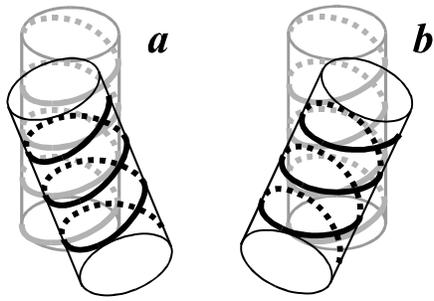

Fig. 9. Molecular alignment at right-handed (*a*) and left-handed (*b*) cholesteric twist. Front molecule is painted black while back molecule is painted gray. Note that right-handed twist allows the front molecule to fit its helical strands in the middle of the groove on the opposing surface. In contrast, left-handed twist results in strands crossing each other. Intuitively, for single-stranded, right-handed helices (such as shown) the right-handed cholesteric twist should be more energetically favorable both for electrostatic and steric reasons. One may expect this trend to be preserved for double-stranded helices such as DNA, but the geometry of interaction between double-stranded helices is more complicated. As a result, an inversion of the torque direction becomes possible.

This occurs while $(\Delta\phi\text{-}g\Delta z)=0$ is still more energetically favorable than random alignment of nearest neighbor molecules. It follows from Eqs. (23), (24) that the random alignment becomes more favorable at $R<R'_*$, where

$$R'_* \approx R_* - \frac{\ln(4)}{(\kappa_2 - \kappa_1)} \ . \tag{34}$$

Since $\ln(\kappa_2/\kappa_1) = 0.5\ln\left[(\kappa^2+4g^2)/(\kappa^2+g^2)\right] < \ln(4)$ we find that $R'_* < \tilde{R}_*$ (Fig. 8a).

Thus, a change in the interlayer torque sign from positive to negative and the corresponding inversion of the cholesteric twist sense from right-handed to left-handed is, in principle, possible in a small interval of interaxial distances. In other words, the twist sense is not uniquely defined by the handedness of the molecules. It may also depend on the cholesteric phase density and on intricate details of charge patterns on molecular surfaces.

This simplified example, however, is intended only to bring up the issue. As discussed above, pairwise alignment of DNA becomes frustrated at $R<R_*$ and it may significantly deviate from $(\Delta\phi\text{-}g\Delta z)=0$ without becoming random. The true prediction of the twist sense within our model requires a much more complicated many-body calculation of frustrated pairwise alignment which is beyond the scope of the present work.

Experimentally, the twist sense of cholesteric aggregates of DNA was studied by circular dichroism (CD) in the absorption band of DNA base pairs (~260 nm) [66]. Anomalously strong, negative CD spectra measured in these experiments were attributed to the left-handed cholesteric twist. Such interpretation is, however, debatable [67]. Nevertheless, if the left-handed twist is not an experimental artifact [68], the possibility of the cholesteric twist sense inversion may provide a clue to explaining this counterintuitive observation.

## VII. CONCLUSIONS

This work is a first step toward a semi-microscopic theory of chiral liquid crystalline phases formed by charged, helical macromolecules. Here we report a closed form, explicit expression for a chiral interaction potential between two rigid helices at arbitrary, fixed mutual orientations. This bare, "ground state" pair potential has a complex landscape (Fig. 3) and a number of nontrivial features. Understanding of these features is essential for developing a future statistical theory that would account for thermal motions and many-body effects. As an illustration of possible manifestations of pair potential features in real phenomena, we attempt to qualitatively rationalize some of the observed properties of cholesteric aggregates of DNA.

Probably the most nontrivial feature of the potential is the following. At small interaxial angles $\psi$ ($|\sin\psi|<<1$), the chiral contribution to the energy of interaction between two *infinitely long* helices depends on $\psi$ as $\sin\psi/|\sin\psi|$, i.e. it depends on the direction but not on the amplitude of the twist. This is not true for the interaction between two helices of *finite length*. The molecular length plays a critical role in the force balance.

Our estimates (Sections VI.E and D) suggest that the effect of molecular length may be the key to understanding the origin of the large pitch $P$~0.4-5 $\mu$m observed in cholesteric aggregates of 150 bp long DNA fragments ($L$~500 Å) [5,6,10]. The interaxial angle between helices in adjacent layers favored by the bare potential corresponds to the cholesteric pitch from $P_{\min} \approx L\sqrt{3\pi R_{\min}\kappa_1/2} \approx 0.3$ $\mu$m to $P_{\max} \sim 10L\sqrt{3\pi R_{\max}\kappa_2/2} \approx 7$ $\mu$m. Here $R_{\min}\approx 32$ Å and $R_{\max}\approx 49$ Å are the minimal and maximal interaxial separations between nearest neighbor helices in the cholesteric phase and $\kappa_1\approx 0.5$ Å$^{-1}$ and $\kappa_2\approx 0.9$ Å$^{-1}$ are defined by Eq. (11).

While conformational and positional disorder might affect intermolecular interactions, the agreement of the pitch estimate based on bare interaction between immobile helices with experimental data is probably not just a coincidence. Molecular motions that are expected to disrupt chiral interactions are rotation about the principal axis and axial translation. However, evaluation of their energetic cost suggests that such motions should have small amplitudes (see Section VI.A). Bending fluctuations are expected to hinder the rotation and axial translation even further without a direct effect on the relative balance between chiral and non-chiral torques. The only unhindered motion in the cholesteric phase seems to be a screw-like axial translation-plus-rotation that does not change the alignment of charged strands on opposing molecules. It could be responsible for the elimination of long-range positional order, but it should not affect intermolecular forces, except for minor "edge effects" associated with translation of molecular tips. Thus, we expect it to have only a minor (if any) effect on the cholesteric pitch. It is, therefore, conceivable that long-range positional disorder may coexist with strong alignment of nearest neighbor molecules whose effective interaction is described by a weakly perturbed ground state potential.

The stability of the nearest neighbor alignment depends on average interaxial separation in the cholestereic phase. At large separations, the energetic cost of about-the-axis rotations and axial translations decreases. Notably, the boundary of the existence of the cholesteric phase ($R\approx 49$ Å) coincides with the separation at which the height of the barriers preventing about-the-axis rotations reduces to 0.5-1 $k_BT$. At small separations, $R<R_*=30$-40 Å (Fig. 8a), optimal alignment of nearest neighbor pairs becomes incompatible with the locally hexagonal symmetry of the cholesteric phase (see Section VI. E). Such alignment frustration is also likely to reduce the cost of about-the-axis rotations. This may explain the loss of chiral interactions at small distances and the cholesteric-hexagonal transition in DNA at $R \approx 32$ Å [12,27].

Finally, the pair potential suggests that the sense of the cholesteric twist may be determined not only by the handedness of molecules in the aggregate, but also by counterion adsorption pattern and by interaxial separation between helices. As a result, an inversion of the cholesteric twist in DNA assemblies (from right-handed at large separations to left-handed at small separations) may occur (Fig. 8b). Such inversion may explain the left-handed twist in DNA aggregates deduced [66] from anomalously strong, negative circular dichroism spectra of polymer-salt-induced DNA aggregates (see Section VI.F).

The agreement between the estimates based on the pair interaction potential and experiments indicates that we may be on the right track to understanding the complex physics of chiral aggregates of charged, helical macromolecules. However, the final verdict requires a rigorous many-body statistical theory which is still to be worked out.

## Acknowledgments

The authors are thankful to Yu.M. Evdokimov, P.-L. Hansen, A.R. Khokhlov, A.J. Liu, T.C. Lubensky, S. Malinin, V.A. Parsegian, R. Podgornik, and D.C. Rau for useful discussions. A part of this work was performed within the 1998 summer-fall program on "Electrostatic Effects in Complex Fluids and Biophysics" at the Institute for Theoretical Physics, University of California at Santa Barbara (ITP). We are thankful to the program organizers: W.M. Gelbart, V.A. Parsegian, and P. Pincus - for the invitation to this program, the ITP staff for hospitality, and acknowledge the National Science Foundation, Grant No. PHY94-07194 which made our participation possible. AAK also appreciates the support of this work by the Deutsche Forschungsgemeinschaft, Grant KO 1391/4-1, and the financial support of his regular visits to Bethesda by the National Institute of Child Health and Human Development, NIH.

## Appendix

## CALCULATION OF PAIR INTERACTION POTENTIAL

### 1. OUTLINE

Consider interaction between two infinitely long, rigid molecules with arbitrary surface charge patterns. The molecules may cross at an arbitrary angle $\psi$. The distance of closest approach between their axes is $R$. The molecules have cylindrical inner cores with the same radius $a$. The molecular cores do not intersect ($R>2a$). We assume that fixed surface charges, adsorbed ions, and condensed counterions lie at the core/water interface (see Section IIA). The molecular configuration and the coordinate systems we use to describe it are shown in Fig. 1.

We describe all charges at each core/water interface explicitly by the charge densities $\rho_v(\mathbf{r})$ defined in Cartesian coordinates of the "laboratory" frame whose $z$-axis coincides with the long axis of molecule 2 (Fig. 1). To calculate the electric field outside inner molecular cores, we use the standard method of induced surface charges.

Specifically, the Fourier transform of the electrostatic potential $\varphi(\mathbf{r})$ is given by

$$\tilde{\varphi}(\mathbf{k}) = \sum_{v=1}^{2} G(k)\left(\tilde{\rho}_v(\mathbf{k}) + \tilde{\rho}_v^{ind}(\mathbf{k})\right). \tag{A 1}$$

Here $\tilde{\rho}_v(\mathbf{k})$ and $\tilde{\rho}_v^{ind}(\mathbf{k})$ are the Fourier transforms of $\rho_v(\mathbf{r})$ and of the induced charge density, respectively.

$$G(k) = \frac{4\pi}{\varepsilon_s \left(k^2 + \kappa^2\right)} \tag{A 2}$$

is the electrostatic Green's function in the **k**-space, $k=|\mathbf{k}|$, $\kappa^{-1}=\lambda_D$ is the Debye length, and $\varepsilon_s$ is the solvent (water) dielectric constant. The Fourier transform of any function $f(\mathbf{r})$ is defined as $\tilde{f}(\mathbf{k}) = (2\pi)^{-3/2} \int d^3\mathbf{r}\, f(\mathbf{r}) e^{i\mathbf{kr}}$.

Because of the cylindrical shape of inner molecular cores, it is convenient to describe the surface charge densities in two separate "molecular" frames in cylindrical coordinate systems $(z, \phi, r)$, each associated with the corresponding molecular axis (see Fig. 1),

$$\rho_v(\mathbf{r}) = \rho_v(z, \phi, r) = \sigma_v(z, \phi)\delta(r - a) \tag{A 3}$$

In the first order approximation, we truncate the summation of consecutive images after the leading term. Then, the electrostatic interaction energy is given by,

$$E_{int} \approx \int d^3\mathbf{k}\left(\tilde{\rho}_1(\mathbf{k}) + \tilde{\rho}_{1,0}^{ind}(\mathbf{k})\right)\left(\tilde{\rho}_2(-\mathbf{k}) + \tilde{\rho}_{2,0}^{ind}(-\mathbf{k})\right)G(k). \tag{A 4}$$

where $\tilde{\rho}_{v,0}^{ind}(\mathbf{k})$ is the density of the induced charge on the surface of the core of the molecule $v$ in the absence of the second molecule. This approximation is appropriate when the surface separation between molecules is larger than the Debye length. By recalculating $\tilde{\rho}_v(\mathbf{k})$ and $\tilde{\rho}_{v,0}^{ind}(\mathbf{k})$ from $\tilde{\sigma}_v(q,n)$ we find an expression for the interaction energy in terms of $\tilde{\sigma}_v(q,n)$.

## 2. FOURIER TRANSFORMS OF CHARGE DENSITY

**Molecule "2"**

The long axis of this molecule coincides with $z$ axis of the chosen Cartesian "laboratory frame" (Fig. 1). We can therefore use the expression derived in our previous work[27]

$$\tilde{\rho}_2(\mathbf{k}) + \tilde{\rho}_2^{ind}(\mathbf{k}) = \frac{a}{(2\pi)^{1/2}} \sum_{m=-\infty}^{\infty} i^m \left(\tilde{\sigma}_2(q,m) + \tilde{\sigma}_2^{ind}(q,m)\right) J_m(Ka) e^{-im\phi_\mathbf{K}} \tag{A 5}$$

This expression converts the Fourier transforms of the charge density from cylindrical coordinates (coaxial with the main axis of the molecule) into the Cartesian coordinates, **k**=$(q,K,\phi_K)$.

While we treat the surface density of fixed and adsorbed charges, $\tilde{\sigma}_2(q,m)$, as known, the density of induced charges, $\tilde{\sigma}_2^{ind}(q,m)$, is to be calculated. When the molecules are not too close to each other, the series of consecutive images rapidly converges. In the first order approximation, we may replace $\tilde{\sigma}_2^{ind}(q,m)$ by $\tilde{\sigma}_{2,0}^{ind}(q,m)$ which is the surface density of induced charges in the absence of molecule 1. To calculate $\tilde{\sigma}_{2,0}^{ind}(q,m)$, we first find the electrostatic potential created by $\sigma_2$ and $\sigma_{2,0}^{ind}$, as if $\sigma_{2,0}^{ind}$ were known. Then, we use the continuity of the potential and of the normal component of electric induction at the inner-core/water interface to find $\sigma_{2,0}^{ind}$.

The electrostatic potential created by molecule 2 in the absence of molecule 1 is

$$\tilde{\varphi}_2(\mathbf{k}) + \tilde{\varphi}_{2,0}^{ind}(\mathbf{k}) = \frac{2a(2\pi)^{1/2}}{\varepsilon_s \left(k^2 + \kappa^2\right)} \sum_{m=-\infty}^{\infty} i^m \left(\tilde{\sigma}_2(q,m) + \tilde{\sigma}_{2,0}^{ind}(q,m)\right) J_m(Ka) e^{-im\phi_K} \quad (A\ 6)$$

In cylindrical coordinates:

$$\varphi_2(z,\phi,r) + \varphi_{2,0}^{ind}(z,\phi,r) = \frac{a}{\pi\varepsilon_s} \int_{-\infty}^{\infty} dq \int_0^{\infty} KdK \int_0^{2\pi} d\phi_K \frac{e^{-iqz} e^{-iKr\cos(\phi_K - \phi_r)}}{\left(q^2 + K^2 + \kappa^2\right)}$$
$$\sum_{m=-\infty}^{\infty} i^m \left(\tilde{\sigma}_2(q,m) + \tilde{\sigma}_{2,0}^{ind}(q,m)\right) J_m(Ka) e^{-im\phi_K} \quad (A\ 7)$$

The integration over $\phi_K$ gives

$$\tilde{\varphi}_2(q,m,r) + \tilde{\varphi}_{2,0}^{ind}(q,m,r) = \frac{4\pi a}{\varepsilon_s} \left(\tilde{\sigma}_2(q,m) + \tilde{\sigma}_{2,0}^{ind}(q,m)\right) \int_0^{\infty} KdK \frac{J_m(Ka) J_m(Kr)}{\left(q^2 + K^2 + \kappa^2\right)} \quad (A\ 8)$$

Note that fixed and adsorbed charges are located at $r=a+0$ while induced charges are located at $r=a-0$. The potential inside this infinitesimally thin layer ($a-0<r<a+0$) is given by

$$\tilde{\varphi}_2(q,m,r) + \tilde{\varphi}_{2,0}^{ind}(q,m,r) = \frac{4\pi a}{\varepsilon_s} \left(\tilde{\sigma}_2(q,m) K_m(\tilde{\kappa}_q a) I_m(\tilde{\kappa}_q r) + \tilde{\sigma}_{2,0}^{ind}(q,m) I_m(\tilde{\kappa}_q a) K_m(\tilde{\kappa}_q r)\right) \quad (A\ 9)$$

where

$$\tilde{\kappa}_q = \sqrt{\kappa^2 + q^2} \quad (A\ 10)$$

The potential inside the molecular core is a solution of the Poisson equation whose general form is

$$\tilde{\varphi}_2^{core}(q,m,r) = B(q,m) I_m(|q|r) \tag{A 11}$$

where $B(q,m)$ is an integration constant. From continuity of the potential at the core we find

$$B(q,m) = \frac{4\pi a}{\varepsilon_s} \left( \tilde{\sigma}_2(q,m) + \tilde{\sigma}_{2,0}^{ind}(q,m) \right) K_m(\tilde{\kappa}_q a) \frac{I_m(\tilde{\kappa}_q a)}{I_m(|q|a)} \tag{A 12}$$

and

$$\tilde{\varphi}_2^{core}(q,m,r) = \frac{4\pi a}{\varepsilon_s} \left( \tilde{\sigma}_2(q,m) + \tilde{\sigma}_{2,0}^{ind}(q,m) \right) K_m(\tilde{\kappa}_q a) \frac{I_m(\tilde{\kappa}_q a)}{I_m(|q|a)} I_m(|q|r) \tag{A 13}$$

The continuity of the normal component of electric induction gives

$$\left[ \frac{\partial \left( \tilde{\varphi}_2(q,m,r) + \tilde{\varphi}_{2,0}^{ind}(q,m,r) \right)}{\partial r} \right]_{r=a} = \frac{\varepsilon_c}{\varepsilon_s} \left[ \frac{\partial \tilde{\varphi}_2^{core}(q,m,r)}{\partial r} \right]_{r=a}, \tag{A 14}$$

where $\varepsilon_c$ and $\varepsilon_s$ are the dielectric constants inside and outside of molecular cores, respectively. After substituting Eqs. (A 9), (A 13) into Eq. (A 14) we obtain[69]

$$\left( \tilde{\sigma}_2(q,m) + \tilde{\sigma}_{2,0}^{ind}(q,m) \right) = -\frac{\tilde{\sigma}_2(q,m)}{\tilde{\kappa}_q a \left( 1 - \tilde{\beta}_m(q) \right) I_m(\tilde{\kappa}_q a) K'_m(\tilde{\kappa}_q a)} \tag{A 15}$$

where

$$\tilde{\beta}_m(q) = \frac{\varepsilon_c}{\varepsilon_s} \frac{|q|}{\tilde{\kappa}_q} \frac{K_m(\tilde{\kappa}_q a) I'_m(|q|a)}{I_m(|q|a) K'_m(\tilde{\kappa}_q a)} \tag{A 16}$$

Finally, after substitution of Eq. (A 16) into Eq. (A 5), we arrive at

$$\tilde{\rho}_2(\mathbf{k}) + \tilde{\rho}_{2,0}^{ind}(\mathbf{k}) = -\frac{1}{(2\pi)^{1/2}} \sum_{m=-\infty}^{\infty} i^m \frac{\tilde{\sigma}_2(q,m) J_m(Ka) e^{-im\phi_K}}{\tilde{\kappa}_q \left( 1 - \tilde{\beta}_m(q) \right) I_m(\tilde{\kappa}_q a) K'_m(\tilde{\kappa}_q a)} \tag{A 17}$$

**Molecule "1"**

The calculation of $\tilde{\rho}_1(\mathbf{k}) + \tilde{\rho}_{1,0}^{ind}(\mathbf{k})$ is more involved because molecule 1 is shifted and rotated with respect to the $z$-axis of our Cartesian laboratory frame. Thus, we introduce two auxiliary Cartesian coordinate systems as follows (Fig. 1):
- $\mathbf{r}' \equiv (x', y', z')$ is shifted from $(x, y, z)$ by the vector $-\mathbf{R}$ connecting the points of the closest approach of the axes of molecules 1 and 2;
- $\mathbf{r}'' \equiv (x'', y'', z'')$ is obtained by rotation of $(x', y', z')$ by the angle $\psi$ so that $z''$ coincides with the axis of molecule 1.

We use $\rho'$ and $\rho''$ to denote the charge density function in $(x', y', z')$ and $(x'', y'', z'')$ coordinates, respectively.

Since

$$\rho_1(\mathbf{r}) + \rho_{1,0}^{ind}(\mathbf{r}) = \rho_1'(\mathbf{r}+\mathbf{R}) + \rho_{1,0}^{ind\,'}(\mathbf{r}+\mathbf{R}), \quad (A\ 18)$$

we find

$$\tilde{\rho}_1(\mathbf{k}) + \tilde{\rho}_{1,0}^{ind}(\mathbf{k}) = e^{-i\mathbf{k}\mathbf{R}}\left(\tilde{\rho}_1'(\mathbf{k}) + \tilde{\rho}_{1,0}^{ind\,'}(\mathbf{k})\right). \quad (A\ 19)$$

Introducing an Euler rotation matrix $\vec{\vec{T}}$,

$$\mathbf{r}'' = \vec{\vec{T}}\mathbf{r}' \quad (A\ 20)$$

$$\vec{\vec{T}} = \begin{pmatrix} 1 & 0 & 0 \\ 0 & \cos\psi & -\sin\psi \\ 0 & \sin\psi & \cos\psi \end{pmatrix} \quad (A\ 21)$$

we obtain

$$\tilde{\rho}'(\mathbf{k}) = \frac{1}{(2\pi)^{3/2}}\int d^3\mathbf{r}''\rho''(\mathbf{r}'')e^{i\mathbf{k}\mathbf{T}^{-1}\mathbf{r}''} = \frac{1}{(2\pi)^{3/2}}\int d^3\mathbf{r}''\rho''(\mathbf{r}'')e^{i\boldsymbol{\mu}\mathbf{r}''} = \tilde{\rho}''(\boldsymbol{\mu}) \quad (A\ 22)$$

where

$$\boldsymbol{\mu} = \vec{\vec{T}}\mathbf{k} = \begin{pmatrix} k_x \\ k_y\cos\psi - k_z\sin\psi \\ k_y\sin\psi + k_z\cos\psi \end{pmatrix} \quad (A\ 23)$$

From Eqs. (A 19) and (A 22) we find

$$\tilde{\rho}_1(\mathbf{k}) + \tilde{\rho}_{1,0}^{ind}(\mathbf{k}) = e^{-i\mathbf{k}\mathbf{R}}\left(\tilde{\rho}_1''(\boldsymbol{\mu}) + \tilde{\rho}_{1,0}^{ind\,''}(\boldsymbol{\mu})\right) \quad (A\ 24)$$

and by analogy with Eq. (A 17) we arrive at

$$\tilde{\rho}_1(\mathbf{k}) + \tilde{\rho}_{1,0}^{ind}(\mathbf{k}) = -\frac{e^{-iKR\cos(\phi_K)}}{(2\pi)^{1/2}} \sum_{n=-\infty}^{\infty} i^n \frac{\tilde{\sigma}_1(\mu_z,n) J_n(Ma) e^{-in\phi_M}}{\tilde{\kappa}_{\mu_z} \left(1 - \tilde{\beta}_n(\mu_z)\right) I_n(\tilde{\kappa}_{\mu_z} a) K'_n(\tilde{\kappa}_{\mu_z} a)} \quad (A\,25)$$

where the relationship between the cylindrical coordinates of the vectors $\mathbf{k}=(q,K,\phi_K)$ and $\vec{\mu}=(\mu_z,M,\phi_M)$ is given by

$$\mu_z = q\cos\psi + K\sin\phi_K \sin\psi, \quad (A\,26)$$

$$M = \sqrt{K^2 + q^2 - \mu_z^2}, \quad (A\,27)$$

$$\cos\phi_M = \frac{K}{M}\cos\phi_K, \qquad \sin\phi_M = \frac{K\sin\phi_K \cos\psi - q\sin\psi}{M}, \quad (A\,28)$$

where the second part of Eq. (A 28) is used to determine the quadrant of $\phi_M$.

## 3. INTERACTION BETWEEN MOLECULES WITH ARBITRARY SURFACE CHARGE PATTERNS

After substituting Eqs. (A 17), (A 25), and (A 2) into Eq. (A 4), we obtain

$$E_{int} \approx \frac{2}{\varepsilon_s} \sum_{n,m=-\infty}^{\infty} i^{n-m} \int_0^{\infty} KdK \int_0^{2\pi} d\phi_K \int_{-\infty}^{\infty} \frac{dq}{\left(K^2 + \tilde{\kappa}_q^2\right)}$$
$$\left( \frac{\tilde{\sigma}_1(\mu_z,n)\tilde{\sigma}_2(-q,-m) J_n(Ma) J_m(Ka) e^{-iKR\cos(\phi_K)} e^{-in\phi_M} e^{im\phi_K}}{\tilde{\kappa}_{\mu_z} \tilde{\kappa}_q \left(1-\tilde{\beta}_n(\mu_z)\right)\left(1-\tilde{\beta}_m(q)\right) K'_n(\tilde{\kappa}_{\mu_z} a) K'_m(\tilde{\kappa}_q a) I_n(\tilde{\kappa}_{\mu_z} a) I_m(\tilde{\kappa}_q a)} \right) \quad (A\,29)$$

This expression relates the interaction Hamiltonian with the surface charge densities on any two crossed molecules regardless of the details of their surface charge patterns. It is applicable to interaction between rods with arbitrary surface charge distributions. This expression is not exact because it is based on the first order approximation for induced charge densities. It becomes exact in nonpolar media where $\varepsilon_s = \varepsilon_c = \varepsilon$ and $\kappa=0$. In this case, it has the form

$$E_{int} = \frac{2a^2}{\varepsilon} \sum_{n,m=-\infty}^{\infty} i^{n-m} \int_0^{\infty} KdK \int_0^{2\pi} d\phi_K \int_{-\infty}^{\infty} \frac{dq}{\left(K^2 + q^2\right)}$$
$$\left( \tilde{\sigma}_1(\mu_z,n)\tilde{\sigma}_2(-q,-m) J_n(Ma) J_m(Ka) e^{-iKR\cos(\phi_K)} e^{-in\phi_M} e^{im\phi_K} \right) \quad (A\,30)$$

# 4. INTERACTION BETWEEN MOLECULES WITH HELICAL SURFACE CHARGE PATTERNS

After substitution of Eq. (6) into Eq. (A 29), we arrive at

$$E_{int} \approx \frac{8\pi^2 \sigma_0^2}{\varepsilon_s} \sum_{n,m=-\infty}^{\infty} \int_0^\infty \frac{KdK}{\left(K^2 + \tilde{\kappa}_q^2\right)} \int_0^{2\pi} d\phi_{\mathbf{K}} \int_{-\infty}^{\infty} dq \left[ e^{-iKR\cos(\phi_{\mathbf{K}})} i^{n-m} e^{-in\phi_{\mathbf{M}}} e^{im\phi_{\mathbf{K}}} e^{in(gz_1-\phi_1)} e^{-im(gz_2-\phi_2)} \right]$$
$$\times \left( \frac{\tilde{p}(\mu_z)\tilde{p}(-q)\delta(\mu_z+ng)\delta(q+mg)J_n(Ma)J_m(Ka)}{\tilde{\kappa}_{\mu_z}\tilde{\kappa}_q\left(1-\tilde{\beta}_n(\mu_z)\right)\left(1-\tilde{\beta}_m(q)\right)K'_n(\tilde{\kappa}_{\mu_z}a)K'_m(\tilde{\kappa}_q a)I_n(\tilde{\kappa}_{\mu_z}a)I_m(\tilde{\kappa}_q a)} \right)$$
(A 31)

The delta function, $\delta(q+mg)$, removes the integral over $q$. Simplifying the resulting expression, we find

$$E_{int} \approx \frac{8\pi^2 \sigma_0^2}{\varepsilon_w |\sin\psi|} \sum_{n,m=-\infty}^{\infty} \int_0^\infty \frac{dK}{\left(K^2 + \kappa_m^2\right)}$$
$$\times \frac{\tilde{p}(-ng)\tilde{p}(mg)J_n(Ma)J_m(Ka)}{\kappa_n \kappa_m \left(1-\tilde{\beta}_n(ng)\right)\left(1-\tilde{\beta}_m(mg)\right)K'_n(\kappa_n a)K'_m(\kappa_m a)I_n(\kappa_n a)I_m(\kappa_m a)}$$
$$\times \int_0^{2\pi} d\phi_{\mathbf{K}} \delta\left(\sin\phi_K + \frac{u_{n,m}(\psi)}{K}\right)$$
$$\times \cos\left[KR\cos(\phi_{\mathbf{K}}) + n\phi_{\mathbf{M}}(\phi_{\mathbf{K}}) - m\phi_{\mathbf{K}} + (m-n)\frac{\pi}{2} - n(gz_1-\phi_1) + m(gz_2-\phi_2)\right]$$
(A 32)

where

$$u_{n,m}(\psi) = \frac{ng - mg\cos\psi}{\sin\psi} \tag{A 33}$$

and

$$\kappa_n = \sqrt{\kappa^2 + n^2 g^2} \tag{A 34}$$

Let us use the equality

$$\delta\left(\sin\phi_{\mathbf{K}} + \frac{u_{n,m}(\psi)}{K}\right) = \frac{\Theta(K - |u_{n,m}(\psi)|)}{|\cos\phi_{\mathbf{K}}|} \left[\delta\left(\phi_{\mathbf{K}} - \phi_{\mathbf{K}}^\#\right) + \delta\left(\phi_{\mathbf{K}} - \pi + \phi_{\mathbf{K}}^\#\right)\right], \tag{A 35}$$

where $\Theta(x)$ is the Heaviside step-function ($\Theta(x)=1$ at $x \geq 0$ and $\Theta(x)=0$ at $x<0$) and

$$\phi_{\mathbf{K}}^\# = \arcsin\left[\frac{-u_{n,m}(\psi)}{K}\right] = \arctan\left[\frac{-u_{n,m}(\psi)}{\sqrt{K^2 - u_{n,m}^2(\psi)}}\right]. \tag{A 36}$$

Using Eq. (A 28) for $\phi_M$ ($\phi_K$), we find

$$\phi_M\left(\phi_K^{\#}\right) = \arctan\left[\frac{u_{m,n}(\psi)}{\sqrt{K^2 - u_{n,m}^2(\psi)}}\right]. \tag{A 37}$$

After we substitute Eqs. (A 35), (A 37) into Eq. (A 32), introduce a new variable

$$t = \sqrt{K^2 - u_{n,m}^2(\psi)}, \tag{A 38}$$

and calculate the integral over $\phi_K$, we obtain

$$E_{\text{int}} \approx \frac{8\pi^2 \sigma_0^2}{\varepsilon_w |\sin\psi|} \sum_{n,m=-\infty}^{\infty} \frac{(-1)^n \tilde{p}(-ng)\tilde{p}(mg)\cos\left[n(\phi_1 - gz_1) - m(\phi_2 - gz_2)\right] \mathbf{I}_{n,m}}{\kappa_n \kappa_m \left(1 - \tilde{\beta}_n(ng)\right)\left(1 - \tilde{\beta}_m(mg)\right) K'_n(\kappa_n a) K'_m(\kappa_m a) I_n(\kappa_n a) I_m(\kappa_m a)}, \tag{A 39}$$

where

$$\mathbf{I}_{n,m} = 2 \int_0^\infty dt \, \frac{J_n\left[a\sqrt{t^2 + u_{m,n}^2(\psi)}\right] J_m\left[a\sqrt{t^2 + u_{n,m}^2(\psi)}\right]}{\left(t^2 + u_{n,m}^2(\psi) + \kappa_m^2\right)}$$
$$\times \cos\left[Rt + n\left(\arctan\left[\frac{u_{m,n}(\psi)}{t}\right] + \frac{\pi}{2}\right) + m\left(\arctan\left[\frac{u_{n,m}(\psi)}{t}\right] + \frac{\pi}{2}\right)\right] \tag{A 40}$$

**Calculation of $\mathbf{I}_{n,m}$ by contour integration**

On the complex plane, the integrand in Eq. (A 40) has a pole at $t = \pm i\sqrt{u_{n,m}^2(\psi) + \kappa_m^2}$ and several branching points. The contributions from the latter can be removed after a set of exact transformations of the integral, which are different for zero and nonzero values of $u_{n,m}(\psi)$.

First, consider the case when both $u_{n,m}(\psi) \neq 0$ and $u_{m,n}(\psi) \neq 0$. In order to remove a branching point at $t=0$, we use the identity

$$\arctan\left(\frac{u_{n,m}(\psi)}{t}\right) = -\arctan\left(\frac{t}{u_{n,m}(\psi)}\right) + \text{sgn}\left(u_{n,m}(\psi)\right)\frac{\pi}{2}, \tag{A 41}$$

where the sign function is defined as $\text{sign}(x)=x/|x|$ at $x \neq 0$ and $\text{sign}(x)=1$ at $x=0$. Then,

$$\mathbf{I}_{n,m} = (-1)^{\frac{n(1+\text{sgn}[u_{m,n}(\psi)]) + m(1+\text{sgn}[u_{n,m}(\psi)])}{2}} \tilde{\mathbf{I}}_{n,m}, \tag{A 42}$$

where

$$\tilde{\mathbf{I}}_{n,m} = \int_{-\infty}^{\infty} \Delta_{n,m}(t)\, dt \tag{A 43}$$

and

$$\Delta_{n,m}(t) = e^{iRt} \frac{J_n\left[a\sqrt{t^2 + u_{m,n}^2(\psi)}\right] J_m\left[a\sqrt{t^2 + u_{n,m}^2(\psi)}\right]}{\left(t^2 + u_{n,m}^2(\psi) + \kappa_m^2\right)} \left\{ \frac{\left[1 - i\dfrac{t}{u_{m,n}(\psi)}\right]^{\frac{n}{2}} \left[1 - i\dfrac{t}{u_{n,m}(\psi)}\right]^{\frac{m}{2}}}{\left[1 + i\dfrac{t}{u_{m,n}(\psi)}\right]^{\frac{n}{2}} \left[1 + i\dfrac{t}{u_{n,m}(\psi)}\right]^{\frac{m}{2}}} \right\} \tag{A 44}$$

In the derivation we used that arctan($x$) is an odd and cos($x$) is an even function of $x$ and took into account that $\arctan(x) = (i/2)\ln\left[(1-ix)/(1+ix)\right]$. The Bessel functions $J_n$ and $J_m$ in $\Delta_{n,m}(t)$ have branching points at $t = \pm i u_{m,n}(\psi)$ that are compensated by the corresponding branching points of the fraction in curly brackets. Thus,

$$\tilde{\mathbf{I}}_{n,m} = 2\pi i \,\mathrm{Res}\{\Delta_{n,m}\}_{t=i\sqrt{u_{n,m}^2(\psi)+\kappa_m^2}}$$

$$= \pi(-1)^{\frac{n[1-\mathrm{sgn}(u_{m,n}(\psi))] + m[1-\mathrm{sgn}(u_{n,m}(\psi))]}{2}} e^{-R\sqrt{u_{n,m}^2(\psi)+\kappa_m^2}} \frac{I_n[a\kappa_n] I_m[a\kappa_m]}{\sqrt{u_{n,m}^2(\psi) + \kappa_m^2}} \tag{A 45}$$

$$\times \left[ \frac{\sqrt{u_{m,n}^2(\psi) + \kappa_n^2} + u_{m,n}(\psi)}{\sqrt{u_{m,n}^2(\psi) + \kappa_n^2} - u_{m,n}(\psi)} \right]^{\frac{n}{2}} \left[ \frac{\sqrt{u_{n,m}^2(\psi) + \kappa_m^2} + u_{m,n}(\psi)}{\sqrt{u_{n,m}^2(\psi) + \kappa_m^2} - u_{m,n}(\psi)} \right]^{\frac{m}{2}}$$

At cos$\psi$=$n/m$, $u_{n,m}(\psi)$=0 while at cos$\psi$=$m/n$, $u_{m,n}(\psi)$=0. Eq. (A 41) cannot be used at $u_{n,m}(\psi)$=0 or $u_{m,n}(\psi)$=0 ($u_{n,m}$ and $u_{m,n}$ cannot have zero values simultaneously at nonzero $\psi$). Consider, e.g., the case when $u_{n,m}(\psi)$=0 and $u_{m,n}(\psi)\neq$0. Then,

$$\mathbf{I}_{n,m} = (-1)^m (-1)^{\frac{n(1+\mathrm{sgn}[u_{m,n}(\psi)])}{2}} \tilde{\mathbf{J}}_{n,m}, \tag{A 46}$$

where

$$\tilde{\mathbf{J}}_{n,m} = \int_{-\infty}^{\infty} dt \, \frac{J_n\left[a\sqrt{t^2 + u_{m,n}^2(\psi)}\right] J_m[at]}{\left(t^2 + u_{n,m}^2(\psi) + \kappa_m^2\right)} \cos\left[Rt - n\arctan\left[\frac{t}{u_{m,n}(\psi)}\right] - m\frac{\pi}{2}\right] \tag{A 47}$$

Here we used that the product of $J_m$ and $\cos$ in the integrand is an even function of $t$ regardless of $m$. After performing the integration in the same way as above, we find that

$$\tilde{\mathbf{J}}_{n,m} = \tilde{\mathbf{I}}_{n,m} \tag{A 48}$$

where $\tilde{\mathbf{I}}_{n,m}$ is still given by Eq. (A 45). Furthermore at $u_{n,m}(\psi)=0$, within our definition of sgn($x$), $m\left(1+\text{sgn}\left[u_{n,m}(\psi)\right]\right)/2 = m$.

Thus, whether $u_{n,m}$ or $u_{m,n}$ have zero values or not, we end up with the same result for $\mathbf{I}_{n,m}$ which is

$$\mathbf{I}_{n,m} = \pi(-1)^n(-1)^m e^{-R\sqrt{u_{n,m}^2(\psi)+\kappa_m^2}} \frac{I_n[a\kappa_n]I_m[a\kappa_m]}{\sqrt{u_{n,m}^2(\psi)+\kappa_m^2}}$$

$$\times \left[\frac{\sqrt{u_{m,n}^2(\psi)+\kappa_n^2}+u_{m,n}(\psi)}{\sqrt{u_{m,n}^2(\psi)+\kappa_n^2}-u_{m,n}(\psi)}\right]^{\frac{n}{2}} \left[\frac{\sqrt{u_{n,m}^2(\psi)+\kappa_m^2}+u_{m,n}(\psi)}{\sqrt{u_{n,m}^2(\psi)+\kappa_m^2}-u_{m,n}(\psi)}\right]^{\frac{m}{2}}. \quad (A\,49)$$

After substitution of Eq. (A 49) into Eq. (A 39) we arrive at the formula for the interaction energy given by Eq. (9).

## 5. INTERACTION BETWEEN TWO HELICES WITH PARALLEL LONG AXES

At $\psi=0$, we find that $M=K$, $\mu_z=q$, $\phi_M=\phi_K$. Then, after integrating over $\phi_K$ and subsequently over $K$, Eq. (A 29) takes the form

$$E_{\text{int}} \approx \frac{4\pi}{\varepsilon_s} \sum_{n,m=-\infty}^{\infty} (-1)^m \int_{-\infty}^{\infty} \frac{\tilde{\sigma}_1(q,n)\tilde{\sigma}_2(-q,-m)K_{n-m}(\tilde{\kappa}_q R)\,dq}{\tilde{\kappa}_q^2\left(1-\tilde{\beta}_n(q)\right)\left(1-\tilde{\beta}_m(q)\right)K_n'(\tilde{\kappa}_q a)K_m'(\tilde{\kappa}_q a)} \quad (A\,50)$$

For infinitely long parallel molecules, the total energy of interaction at $\psi=0$ is infinite and the integral in Eq. (A 50) diverges. In this case, the meaningful value is the energy of interaction per unit length, which can be calculated as $\lim_{L\to\infty}(E_{\text{int}}/L)$, where $L$ is the length of the molecules. Then, it is convenient to introduce the charge density pair correlation function

$$s_{1,2}(q,n,m) = \lim_{L\to\infty}\left\{\frac{\tilde{\sigma}_1(q,n)\tilde{\sigma}_2(-q,-m)+\tilde{\sigma}_1(-q,-n)\tilde{\sigma}_2(q,m)}{2L}\right\}. \quad (A\,51)$$

Within our definition of the Fourier transform, $\tilde{\sigma}_1(q,n)\tilde{\sigma}_2(-q,-m)/L$ does not depend on $L$ at $L\to\infty$ and $\int_{-\infty}^{\infty} s_{1,2}(q,n,m)dq$ has a finite value [27]. Then, the energy per unit length of the molecules is given by

$$\frac{E_{\text{int}}}{L} \approx \frac{4\pi}{\varepsilon_s} \sum_{n,m=-\infty}^{\infty} (-1)^m \int_{-\infty}^{\infty} \frac{s_{1,2}(q,n,m)K_{n-m}(\tilde{\kappa}_q R)\,dq}{\tilde{\kappa}_q^2\left(1-\tilde{\beta}_n(q)\right)\left(1-\tilde{\beta}_m(q)\right)K_n'(\tilde{\kappa}_q a)K_m'(\tilde{\kappa}_q a)} \quad (A\,52)$$

For $\tilde{\beta}_n(q)\ll 1$, this expression reduces to our previous result [27], derived for $\varepsilon_c/\varepsilon_s\ll 1$.